\documentclass[cernpreprint, atlasdraft=false, backref=false, coverpage=false, texlive=2023, UKenglish, orcidlogo=true, texmf, orcidlogo]{atlasdoc}
 
\usepackage{atlaspackage}
 
\usepackage{atlasbiblatex}
 
\usepackage[BSM=true, misc=true, process=true, hepprocess=true]{atlasphysics}

\graphicspath{{logos/}{figures/}}
 
\usepackage{ANA-SUSY-2021-06-PAPER-defs}
\usepackage{multirow}

\AtlasTitle{Measurements of $W^{+}W^{-}$ production in decay topologies inspired by searches for electroweak supersymmetry}

\AtlasAbstract{
This paper presents a measurement of fiducial and differential cross-sections for $W^{+}W^{-}$ production in proton--proton collisions at $\sqrt{s}=13$~TeV with the ATLAS experiment at the Large Hadron Collider using a dataset corresponding to an integrated luminosity of 139~fb$^{-1}$. Events with exactly one electron, one muon and no hadronic jets are studied. The fiducial region in which the measurements are performed is inspired by searches for the electroweak production of supersymmetric charginos decaying to two-lepton final states. The selected events have moderate values of missing transverse momentum and the `stransverse mass' variable $m_{\textrm{T2}}$, which is widely used in searches for supersymmetry at the LHC. The ranges of these variables are chosen so that the acceptance is enhanced for direct  $W^{+}W^{-}$ production and suppressed for production via top quarks, which is treated as a background. The fiducial cross-section and particle-level differential cross-sections for six variables are measured and compared with two theoretical SM predictions from perturbative QCD calculations.}

\AtlasRefCode{SUSY-2021-06}
 
\PreprintIdNumber{CERN-EP-2022-087}

\AtlasJournalRef{Eur. Phys. J. C 83 (2023) 718}
\AtlasDOI{10.1140/epjc/s10052-023-11508-9}

\hypersetup{pdftitle={ATLAS document},pdfauthor={The ATLAS Collaboration}}
 
\begin{document}
 
\maketitle

\section{Introduction}
\label{sec:intro}
 
Measurements of $W^{+}W^{-}$ (referred to hereafter as $WW$) production provide important tests of the electroweak (EWK) gauge structure of the Standard Model (SM) of particle physics, and $WW$ production is also an important background process in searches for physics beyond the SM (BSM physics). In searches for supersymmetry~\cite{Golfand:1971iw,Volkov:1973ix,Wess:1974tw,Wess:1974jb,Ferrara:1974pu,Salam:1974ig} (SUSY) where $WW$ events are a significant background, a semi-data-driven approach is often taken, that involves normalising the simulated Monte Carlo (MC) samples to data in a control region (CR), designed to be kinematically similar to the search regions but enriched in SM $WW$ production. Significant deviations of these scaling factors from unity suggest mismodelling in the phase space targeted by the search, but it can be difficult to make comparisons with
the level of agreement observed in precision SM measurements because the scaling factors refer to detector-level quantities which are subject to mis-measurement and inefficiency. Producing "unfolded" particle-level measurements, which are corrected for these effects and can directly be compared with the prediction of a MC event generator, in event topologies associated with search results is a novel way to address this whilst simultaneously extending the programme of precision SM measurements at the LHC. The ATLAS experiment~\cite{PERF-2007-01} has previously reported differential measurements of $t\bar{t}$ and $Z$+jets production in regions related to a search for leptoquarks in dilepton+dijet events~\cite{EXOT-2016-08}. This paper presents the first effort to measure $WW$ production cross-sections in topologies associated with SUSY searches.
 
Inclusive and fiducial $WW$ production cross-sections have been measured in proton--proton ($pp$) collisions at $\sqrt{s}=7$~\TeV~\cite{STDM-2012-01,CMS-SMP-12-005}, 8~\TeV~\cite{STDM-2013-07,CMS-SMP-14-016,STDM-2015-24} and 13~\TeV~\cite{STDM-2015-20,STDM-2017-24,STDM-2018-34,CMS-SMP-18-004} at the LHC, as well as in $e^+e^-$ collisions at LEP~\cite{schael:2013ita} and in $p\bar{p}$ collisions at the Tevatron~\cite{Abe:1996dw,Aaltonen:2009aa,Abazov:2009ys}. This analysis complements existing ATLAS measurements of $WW$ production at 13~\TeV\ in 0-jet events~\cite{STDM-2017-24} and in $\geq 1$-jet events~\cite{STDM-2018-34} by measuring differential cross-sections in a fiducial region kinematically close to the $WW$ control region used in a previous search for electroweak production of supersymmetric charginos or sleptons~\cite{SUSY-2018-32}. That search targeted electroweak production of SUSY particles decaying into final states with two leptons (electrons or muons) and missing transverse momentum using 139~fb$^{-1}$ of $pp$ collisions at 13~\TeV\ collected during Run~2 of the LHC and is referred to hereafter as the `EWK 2$\ell$+0-jets search'.  In that search, $WW$ production was the main background process and the associated theoretical uncertainties were among the dominant systematic uncertainties in the search regions. The present measurement targets event topologies with higher values of the dilepton invariant mass, $m_{e\mu}$, and the magnitude of the missing transverse momentum, \MET, than were used in previous measurements, and can thus be used to provide additional constraints on BSM physics, and probe the expected SM backgrounds for future searches.
 
The $WW \rightarrow e^{\pm}\nu\mu^{\mp}\nu$ decay channel is studied in events with no identified jets with a transverse momentum $p_{\textup{T}}>20$~\GeV\ and pseudorapidity $|\eta|<2.4$,\footnote{ATLAS uses a right-handed coordinate system with its origin at the nominal interaction point (IP) in the centre of the detector and the $z$-axis along the beam pipe. The $x$-axis points from the IP to the centre of the LHC ring, and the $y$-axis points upward. Cylindrical coordinates $(r,\phi)$ are used in the transverse plane, $\phi$ being the azimuthal angle around the $z$-axis. The pseudorapidity is defined in terms of the polar angle $\theta$ as $\eta=-\ln\tan(\theta/2)$} and with \MET between 60 and 80~\GeV. Missing transverse momentum is calculated so as to represent the momentum imbalance in the plane transverse to the colliding beams. High values of \MET\ can be produced when weakly interacting neutral particles escape the detector unseen, and \MET\ is thus an important variable in many BSM searches. This analysis also imposes requirements on the dilepton invariant mass that are more stringent than those in the 36~fb$^{-1}$ $WW$+0-jet measurement~\cite{STDM-2017-24}. The dominant background process is top-quark production (\ttbar\ and single-top $Wt$), which is estimated using the same data-driven method as was used in the EWK 2$\ell$+0-jets search. The measurements are performed in a fiducial phase space close to the geometric and kinematic acceptance of the experimental analysis. Differential cross-section measurements are performed for six variables, which are the same as those considered in the 36~fb$^{-1}$ $WW$+0-jet measurement~\cite{STDM-2017-24}:
 
\begin{itemize}
\item The rapidity of the dilepton system, $|y_{e\mu}|$.
\item The azimuthal separation between the two leptons, $|\Delta\phi_{e\mu}|$.
\item The angular variable $\cos\theta^{\ast}=|\tanh(\Delta y(e\mu)/2)|$, which is longitudinally boost invariant and sensitive to the spin structure of the produced dileptons~\cite{Barr:2005dz}, and where $\Delta y(e\mu)$ is the difference between the electron and muon rapidities.
\item The transverse momentum of the leading lepton, $p_{\mathrm{T}}^{\mathrm{lead}\, \ell}$.
\item The invariant mass of the dilepton system, $m_{e\mu}$.
\item The transverse momentum of the dilepton system, $p_{\mathrm{T}}^{e\mu}$.
\end{itemize}
In this paper, $|y_{e\mu}|$, $|\Delta\phi_{e\mu}|$ and $\cos\theta^{\ast}$ are referred to collectively as `angular' variables, as they probe angular correlations and are sensitive to the spin structure of the $WW$ production system, and $p_{\mathrm{T}}^{\mathrm{lead}\, \ell}$, $m_{e\mu}$ and $p_{\mathrm{T}}^{e\mu}$ are referred to collectively as `scale' variables, as they characterise the energy scale of the process.
 
The rest of this paper is structured as follows. First, Section~\ref{sec:detector} describes the ATLAS detector and then Section~\ref{sec:analysis} presents the  analysis that is performed to measure the fiducial and differential cross-sections. This includes the data and MC samples used, the reconstructed-object definitions and event selections used to define the detector-level signal regions, and the SM background estimation, as well as the systematic uncertainties considered and the unfolding techniques used to correct detector-level information back to particle level. Finally, the results are reported in Section~\ref{sec:results}, and Section~\ref{sec:conclusions} presents the conclusions.
 
\section{ATLAS detector}
\label{sec:detector}
 
The ATLAS experiment at the LHC is a multipurpose particle detector
with a forward--backward symmetric cylindrical geometry and a near \(4\pi\) coverage in
solid angle.
It consists of an inner tracking detector surrounded by a thin superconducting solenoid
providing a \SI{2}{\tesla} axial magnetic field, electromagnetic and hadron calorimeters, and a muon spectrometer.
The inner tracking detector covers the pseudorapidity range \(|\eta| < 2.5\).
It consists of silicon pixel, silicon microstrip, and transition radiation tracking detectors.
Lead/liquid-argon (LAr) sampling calorimeters provide electromagnetic (EM) energy measurements
with high granularity.
A steel/scintillator-tile hadron calorimeter covers the central pseudorapidity range (\(|\eta| < 1.7\)).
The endcap and forward regions are instrumented with LAr calorimeters
for both the EM and hadronic energy measurements up to \(|\eta| = 4.9\).
The muon spectrometer surrounds the calorimeters and is based on
three large superconducting air-core toroidal magnets with eight coils each.
The field integral of the toroids ranges between \num{2.0} and \SI{6.0}{\tesla\metre}
across most of the detector.
The muon spectrometer includes a system of precision tracking chambers and fast detectors for triggering.
A two-level trigger system is used to select events.
The first-level trigger is implemented in hardware and uses a subset of the detector information
to accept events at a rate below \SI{100}{\kHz}.
This is followed by a software-based trigger that
reduces the accepted event rate to \SI{1}{\kHz} on average
depending on the data-taking conditions.
An extensive software suite~\cite{ATL-SOFT-PUB-2021-001} is used in the reconstruction and analysis of real
and simulated data, in detector operations, and in the trigger and data acquisition systems of the experiment.

\section{Analysis}
\label{sec:analysis}
\subsection{Data and simulated event samples}
\label{sec:samples}
 
This analysis uses $pp$ collision data at a centre-of-mass energy of $\sqrt{s} = 13$~\TeV\ collected by the ATLAS detector during the second data-taking run of the LHC, which took place between 2015 and 2018. After applying standard data-quality requirements for LHC and detector operations~\cite{DAPR-2018-01}, this dataset corresponds to a total integrated luminosity of 139~fb$^{-1}$ with an uncertainty of 1.7\% \cite{ATLAS-CONF-2019-021}, obtained using the LUCID-2 sub-detector~\cite{LUCID2} for the primary luminosity measurements. Candidate events were selected by a trigger that required at least one electron--muon pair~\cite{TRIG-2018-05,TRIG-2018-01}. The trigger-level thresholds for the \pT\ of the leptons were 17~\GeV\ for the electron and 14~\GeV\ for the muon. The thresholds applied in the lepton offline selection ensured that trigger efficiencies are constant in the relevant phase space.

Simulated MC event samples are used for the SM background estimates and to correct the signal distributions for detector effects. These were processed through a full simulation of the ATLAS detector~\cite{SOFT-2010-01} based on \GEANT~\cite{Agostinelli:2002hh} and reconstructed with the same algorithms as those used for the data. The generation of the simulated event samples includes the effect of multiple $pp$ interactions per bunch crossing (pile-up), as well as changes in detector response because of interactions in bunch crossings before or after the one containing the hard interaction. Differences between data and simulation in the lepton reconstruction efficiency, energy scale, energy resolution and trigger efficiency~\cite{EGAM-2018-01,PERF-2015-10}, and in the $b$-tagging efficiency~\cite{FTAG-2018-01}, are treated through correction factors that are derived from data and applied as weights to the simulated events. The MC samples are also reweighted so that the distribution of the average number of interactions per bunch crossing reproduces the observed distribution in the data.
 
Simulated $WW$ signal samples were produced by
summing $q\bar{q}$- and $gg$-initiated samples. The $q\bar{q}$-initiated $WW$ signal was simulated at next-to-leading-order (NLO) accuracy in QCD using the \POWHEGBOX[v2]~\cite{Nason:2004rx,Frixione:2007vw,Alioli:2010} generator interfaced to \PYTHIA[8.186]~\cite{Sjostrand:2007gs} for the modelling of the parton shower, hadronisation, and underlying
event, with parameter values set according to the \AZNLO
tune~\cite{STDM-2012-23}. The \CT[10nlo] parton distribution function (PDF) set~\cite{Lai:2010vv} was used
for the hard-scattering processes, whereas the CTEQ6L1 PDF
set~\cite{Pumplin:2002vw} was used for the parton shower~\cite{ATL-PHYS-PUB-2017-005}. The events were normalised to the cross-section calculated to next-to-next-to-leading order (NNLO) in QCD~\cite{Gehrmann:2014fva}.  Loop-induced $gg\rightarrow WW \rightarrow \ell\nu\ell\nu$ events were simulated at LO with up to one
additional parton emission using \SHERPA[2.2.2], with virtual QCD corrections provided by the
\OPENLOOPS library~\cite{Buccioni:2019sur,Cascioli:2011va,Denner:2016kdg,ATL-PHYS-PUB-2017-005}. The $gg \rightarrow WW$ process was normalised to its inclusive NLO QCD cross-section~\cite{Caola:2015rqy}. 
An alternative sample of $q\bar{q} \rightarrow WW$ events was simulated using \SHERPA[2.2.2]~\cite{ATL-PHYS-PUB-2017-005,Bothmann:2019yzt} with matrix elements at NLO accuracy in QCD for up to one additional parton emission and at LO accuracy for up to three additional parton
emissions. For the \SHERPA\ $q\bar{q}$- and $gg$-initiated samples the \NNPDF[3.0nnlo] set of PDFs was used~\cite{Ball:2014uwa}, along with the
dedicated set of tuned parton-shower parameters developed by the
\SHERPA authors. No alternative simulation is considered for the $gg\rightarrow WW$ process, which contributes only a small fraction of the signal.
 
Table~\ref{tab:mc_backgr} summarises the generators used for the SM backgrounds along with the relevant PDF sets, the set of tuned parameters used to configure the hadronisation and underlying event, and the cross-section 
order in \alphas used to normalise the event yields for these samples. This study uses the same simulated samples and groupings for the SM background processes as the EWK 2$\ell$+0-jets search~\cite{SUSY-2018-32}.  The `Others' category groups together processes that produce small or negligible contributions to the signal regions of the search, and includes Drell--Yan, \ttbar\!+$V$ and Higgs boson production.
Further information about the simulations of \ttbar, single-top ($Wt$), multiboson and boson-plus-jet processes can also be found in the relevant public ATLAS notes~\cite{ATL-PHYS-PUB-2018-009,ATL-PHYS-PUB-2016-004,ATL-PHYS-PUB-2017-005,ATL-PHYS-PUB-2017-006}.
\begin{table}[!h]
\tiny
\centering
\caption{\label{tab:mc_backgr}Simulated background event samples with the corresponding matrix element and parton shower (PS) generators, cross-section order in \alphas used to normalise the event yield, underlying-event tune and the generator PDF sets used. Where used, the label `$V$' refers to a $W$ or $Z$ boson.}

\begin{tabular}{lp{0.164\linewidth}p{0.083\linewidth}p{0.132\linewidth}p{0.103\linewidth}p{0.105\linewidth}p{0.090\linewidth}}
\toprule
 
Physics process          			& Generator 	& Parton shower 	&  Normalisation 		& Tune 			& PDF (generator) & PDF (PS)\\
\midrule
 
$t\bar{t}$ 				& \POWHEGBOX[v2]~\cite{Frixione:2007,Frixione:2007vw,Alioli:2010,Campbell:2015}       		& \PYTHIA[8.230]~\cite{Sjostrand:2014zea}  		& NNLO+NNLL~\cite{Czakon:2011xx} 		& A14~\cite{ATL-PHYS-PUB-2014-021} & \NNPDF[3.0nlo]~\cite{Ball:2014uwa} & \NNPDF[2.3lo]~\cite{Ball:2012cx}\\
\midrule
Single top ($Wt$) 				& \POWHEGBOX[v2]~\cite{Frixione:2007vw,Alioli:2010,Re:2011}  	& \PYTHIA[8.230]		&NLO+NNLL~\cite{Aliev:2010zk,Kant:2014oha}		& A14 			& \NNPDF[3.0nlo]~\cite{Ball:2014uwa}
& \NNPDF[2.3lo]
\\
\midrule
 
$VZ  = WZ,ZZ$				& \POWHEGBOX[v2]~\cite{Frixione:2007vw,Alioli:2010,Melia:2011tj,Nason:2013ydw} 		& \PYTHIA[8.210]
& NLO~\cite{ATL-PHYS-PUB-2017-005,Melia:2011tj,Nason:2013ydw}   		& AZNLO~\cite{STDM-2012-23}
&  \CT[10nlo]~\cite{Lai:2010vv} & CTEQ6L1\cite{Pumplin:2002} \\

\midrule
`Others':     &		&		&		&		&		&  \\
\quad Higgs   				& \POWHEGBOX[v2]~\cite{Nason:2004rx, Frixione:2007vw,Alioli:2010} &  \PYTHIA[8.212]~\cite{Sjostrand:2014zea}	       &   NNNLO+NNLL~\cite{deFlorian:2016spz,Anastasiou:2016cez,Anastasiou:2015ema,Dulat:2018rbf,Aglietti:2004nj,Actis:2008ug,Bonetti:2018ukf} &    AZNLO~\cite{STDM-2012-23}	&  \PDFforLHC[15nnlo]~\cite{Butterworth:2015oua}
& CTEQ6L1
\\
\quad  $VVV$				& \SHERPA[2.2.2]~\cite{Gleisberg:2009,Bothmann:2019yzt,ATL-PHYS-PUB-2017-005} 			&  \SHERPA[2.2.2]		& NLO~\cite{Bothmann:2019yzt,ATL-PHYS-PUB-2017-005}			& \SHERPA default~\cite{ATL-PHYS-PUB-2017-005} 	& \NNPDF[3.0nnlo]~\cite{Ball:2014uwa}
& \NNPDF[3.0nnlo]
\\
\quad $t\bar{t}+H$ 			& \MGNLO~\cite{Alwall:2014hca}	& \PYTHIA[8.230]~\cite{Sjostrand:2014zea}					& NLO~\cite{deFlorian:2016spz}			& A14	& \NNPDF[3.0nlo]
& \NNPDF[2.3lo]
\\
 
\quad $t\bar{t}+V$ 			& \MGNLO 			& \PYTHIA[8.210]~\cite{Sjostrand:2014zea}						& NLO~\cite{ATL-PHYS-PUB-2016-005,Alwall:2014hca} 			& A14 & \NNPDF[3.0nlo]
& \NNPDF[2.3lo]
\\
\quad $t\bar{t}+WW$				& \MGNLO			& \PYTHIA[8.186]~\cite{Sjostrand:2007gs} 		& NLO~\cite{Alwall:2014hca}			& A14			& \NNPDF[2.3lo]
&  \NNPDF[2.3lo]
\\  
\quad $tZ,t\bar{t}t\bar{t},t\bar{t}t$		& \MGNLO			& \PYTHIA[8.230]
& NLO~\cite{Alwall:2014hca}      		& A14			& \NNPDF[3.0nlo]
& \NNPDF[2.3lo]
\\
\quad $Z/\gamma(\rightarrow \ell\ell)$+jets 	& \SHERPA[2.2.1]~\cite{Gleisberg:2009,Bothmann:2019yzt,ATL-PHYS-PUB-2016-003}			& \SHERPA[2.2.1]		& NNLO~\cite{Gavin:2010az}			& \SHERPA default~\cite{ATL-PHYS-PUB-2016-003}	& \NNPDF[3.0nnlo]
&\NNPDF[3.0nnlo]
\\
\bottomrule

\end{tabular}
 
\end{table}

\subsection{Event reconstruction and selection}
Events are required to have at least one reconstructed vertex with at least two associated tracks with $\pT >400$~\MeV. When more than one vertex is reconstructed, the one with the highest $\sum\pT^{2}$ of associated tracks is taken to be the primary vertex. All final-state objects (electrons, muons and jets in this study) are required to satisfy `baseline' criteria to ensure they are well-reconstructed and originate from the primary vertex, and additional `signal' criteria are applied to define the objects used in the measurement.
Baseline electrons are required to have $\pT >10$~\GeV, pseudorapidity $|\eta|< 2.47$and a longitudinal impact parameter $z_0$, relative to the primary vertex, satisfying $|z_0 \sin\theta| < 0.5$~mm; baseline muons must fulfill the same criteria except $|\eta|<2.6$. Electrons must satisfy a \textit{Loose} likelihood-based identification requirement~\cite{EGAM-2018-01}, while muons must satisfy the \textit{Medium} identification requirements defined in Ref.~\cite{PERF-2015-10}.  Signal electrons are required to satisfy a \textit{Tight} identification requirement~\cite{EGAM-2018-01} and the track associated with the signal electron is required to have $\vert d_0\vert/\sigma(d_0) < 5$, where $d_0$ is  the transverse impact parameter relative to the primary vertex and $\sigma(d_0)$ is its uncertainty, whilst for signal muons the associated track must have $\vert d_0\vert/\sigma(d_0) < 3$. The  signal-lepton isolation criteria used in the EWK 2$\ell$+0-jets search~\cite{SUSY-2018-32} are also applied in this study.
Hadronic jets are reconstructed from energy deposits in topological clusters of calorimeter cells~\cite{PERF-2014-07,JETM-2018-05} using the anti-$k_t$ algorithm~\cite{Cacciari:2008}, as implemented in the FastJet package~\cite{Fastjet}, with a radius parameter $R=0.4$. They are then calibrated by the application of a jet energy scale derived from 13~\TeV\ data and simulation~\cite{PERF-2016-04}. To reduce the effects of pile-up, for jets with $|\eta|< 2.5$ and $\pt<\SI{120}{\GeV}$ a significant fraction of the tracks associated with each jet must have an origin compatible with the primary vertex, as defined by the jet vertex tagger~\cite{ATLAS-CONF-2014-018}. For jets with $|\eta|>2.5$ and $\pt<\SI{60}{\GeV}$, similar pile-up suppression is achieved through the forward jet vertex tagger~\cite{ATL-PHYS-PUB-2015-034}. Finally, events are rejected if they contain a jet that does not satisfy the jet-quality requirements~\cite{ATLAS-CONF-2015-029,PERF-2014-03}; this removes events impacted by detector noise or non-collision backgrounds. Jets containing $b$-hadrons (`$b$-jets') are identified by the  MV2c10 boosted decision tree algorithm
~\cite{FTAG-2018-01}, using quantities such as the impact parameters of associated tracks along with well-reconstructed secondary vertices. A selection that provides 85\% efficiency for tagging $b$-jets in simulated $t\bar{t}$ events is used in this study.
Only jet candidates with $\pT>\SI{20}{\GeV}$ and $|\eta|<2.4$ are considered,\footnote{Hadronic $\tau$-lepton decay products are treated as jets.} although all jets with $|\eta|<4.9$ are included in the calculation of missing transverse momentum and in the procedure to remove reconstruction ambiguities that could lead to double counting of baseline objects. This procedure is applied as follows:
\begin{itemize}
\item jet candidates within $\Delta R' =\sqrt{(\Delta y)^2+(\Delta\phi)^2} = 0.2$  of an electron candidate are removed; 
\item jets with fewer than three tracks that lie within $\Delta R' = 0.4$ of a muon candidate are removed;
\item electrons and muons within $\Delta R' = 0.4$ of the remaining jets are discarded, to reject leptons from the decay of $b$- or $c$-hadrons;
\item electron candidates 
are rejected if they are found to share an inner-detector track with a muon.
\end{itemize}
The measurements are performed in events with exactly one signal electron and one signal muon with opposite electric charge and each satisfying \pT$>25$~\GeV, and a veto on additional baseline leptons and hadronic jets. 
The multiplicities of non-$b$-tagged jets and $b$-tagged jets are considered separately in the background estimation for this study: events with exactly one $b$-tagged jet with a veto on additional non-$b$-tagged jets are used to estimate and validate the top-quark background.
Requirements are also placed on the missing transverse momentum ($\mathbf p^\mathrm{miss}_\mathrm{T}$), which has magnitude 
\MET. This is defined as the negative vector sum of the transverse momenta of all identified physics objects (electrons, photons, muons and jets), plus an additional `soft term' to include low-momentum tracks associated with the primary vertex but not with these physics objects. The \MET value is adjusted for the calibration of the selected physics objects~\cite{PERF-2016-07}.
 
To access a region of phase space similar to the $WW$ control region in the EWK 2$\ell$+0-jets search, additional requirements are placed on the following variables in this study:
\begin{itemize}
\item The invariant mass of the dilepton system, $m_{e\mu}>100$~\GeV.
\item The magnitude of the missing transverse momentum vector, $\MET \in [60, 80]$~\GeV.
\item The `stransverse mass' variable, $\mttwo \in [60, 80]$~\GeV~\cite{Lester:1999tx,Barr:2003rg}, with \mttwo\ defined as:
 
\begin{equation*}
\mttwo
( \mathbf p_{\mathrm{T,1}}, \mathbf p_{\mathrm{T,2}},\mathbf p^\mathrm{miss}_\mathrm{T}) = \min_{\mathbf q_{\mathrm{T,1}} + \mathbf q_{\mathrm{T,2}} =
\mathbf p^\mathrm{miss}_\mathrm{T} }
\left\{ \max[\;
m_{\mathrm{T}}( \mathbf p_{\mathrm{T,1}}, \mathbf q_{\mathrm{T,1}} ),
m_{\mathrm{T}}( \mathbf p_{\mathrm{T,2}}, \mathbf q_{\mathrm{T,2}}
)
\;]
\right\},
\label{eq:mt2}
\end{equation*}
where $m_{\mathrm T}$ is the transverse mass defined as $m_\textup{T}=\sqrt{2\times|\mathbf p_{\mathrm{T,a}}|\times|\mathbf p_{\mathrm{T,b}}|\times(1-\mathrm{cos}(\Delta\phi))}$, and $\Delta\phi$ is the azimuthal angle between the particles with transverse momenta $\mathbf p_{\mathrm{T,a}}$ and $\mathbf p_{\mathrm{T,b}}$. The vectors $\mathbf p_{\mathrm{T,1}}$ and $\mathbf p_{\mathrm{T,2}}$ are the transverse momenta of the two leptons, and $\mathbf
q_{\mathrm{T,1}}$ and $\mathbf q_{\mathrm{T,2}}$ satisfy
$\mathbf p^\mathrm{miss}_\mathrm{T}= \mathbf q_{\mathrm{T,1}} + \mathbf q_{\mathrm{T,2}}$. The \mttwo\ variable was designed to be sensitive to the mass scales of pair-produced heavy particles that each decay semi-invisibly.  The minimisation is performed over all the possible decompositions of $\mathbf p^\mathrm{miss}_\mathrm{T}$ into two hypothetical invisible particles with momenta $\mathbf
q_{\mathrm{T,1}}$  and  $\mathbf
q_{\mathrm{T,2}}$.
For $t\bar{t}$ or $WW$ decays, assuming an ideal detector with perfect momentum resolution, $m_\textup{T2}( \mathbf p_{\mathrm{T,\ell_1}}, \mathbf p_{\mathrm{T,\ell_2}},\mathbf p^\mathrm{miss}_\mathrm{T})$
has a kinematic endpoint at the mass of the $W$~boson~\cite{Barr:2003rg}. The signal regions of the EWK 2$\ell$+0-jets search required higher values, $\mttwo >100$~\GeV.
\end{itemize}
 
In the EWK 2$\ell$+0-jets search, the top-quark contamination in events with a jet veto was observed to increase with \mttwo\ in the region $\mttwo \in [60, 100]$~\GeV. To maximise $WW$ purity, the control region required $\mttwo \in [60, 65]$~\GeV\ and \MET$\in [60, 100]$~\GeV\ with validation of the estimate being performed in events with $\mttwo \in [65, 100]$~\GeV\ and \MET$>60$~\GeV.
Since $\mttwo$ is sensitive to the angular separation of the lepton pair, the \mttwo\ range is widened for the analysis described in this paper to provide a broader phase space for measuring angular distributions. Since \MET\ and \mttwo\ are correlated, the \MET\ range is tightened to reduce the top-quark contamination. These changes increase the number of events in the region used to perform the differential cross-section measurements without reducing the $WW$ purity.
The previously used requirement on the `object-based \MET\ significance'~\cite{ATLAS-CONF-2018-038} is removed to simplify the definition of the fiducial region at particle level.\footnote{The `object-based \MET significance' helps to separate events with true \MET (arising from weakly interacting particles) from those where it is consistent with particle mismeasurement, resolution effects or identification inefficiencies. On an event-by-event basis, given the full event composition, the \MET significance evaluates the $p$-value that the observed \MET is consistent with the null hypothesis of zero real \MET, as further detailed in Ref.~\cite{ATLAS-CONF-2018-038}.}
The definition of the signal region used for this measurement is summarised in Table~\ref{tab:unfoldingregions}. The same selections are used at particle level when defining the fiducial region used for the fiducial and differential cross-section calculations, as discussed in Section~\ref{sec:unfolding}.
 
\begin{table}[!h]
\caption{\label{tab:unfoldingregions}
Summary of the selection criteria used for the signal region in this study. The same selections are used at detector level and particle level.
}
\small
\begin{center}
\begin{tabular}{l|c}
\hline
Selection requirement & Criteria \\
\hline
Lepton flavour              	 			& $e^{\pm}\mu^{\mp}$   \\
Lepton \pT             	 			& $>25$~\GeV    \\
Lepton $|\eta|$                	 			& $<2.47\,(e^{\pm})$, $<2.6\,(\mu^{\mp})$    \\
Lepton veto              	 			& No additional electrons with $\pT>10$~\GeV, $|\eta|<2.47$    \\
& No additional muons with $\pT>10$~\GeV, $|\eta|<2.6$     \\
 
$m_{e\mu}$                	 			&   $>100$~\GeV   \\
\hline
 
Jet veto	& No jets with $\pT>20$~\GeV, $|\eta|<2.4$			\\
$\mttwo$ 		         			&$\in$ [60,\,80]~\GeV	 \\
$\MET$	        					&$\in$ [60,\,80]~\GeV	 \\
\hline
\end{tabular}
\end{center}
\end{table}
 
\begin{figure}[htb!]
\centering
\includegraphics[width=0.43\linewidth]{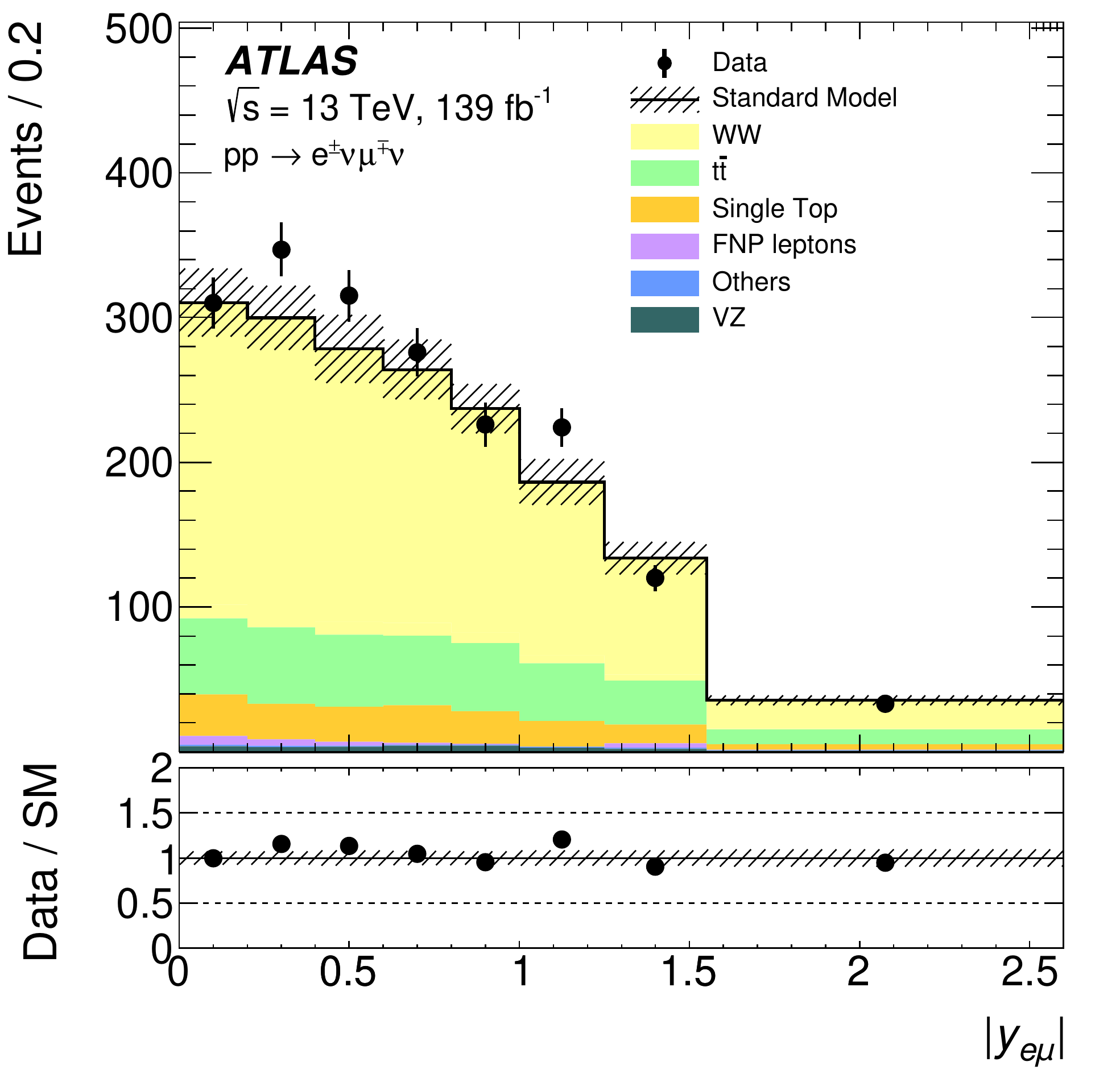}
\includegraphics[width=0.43\linewidth]{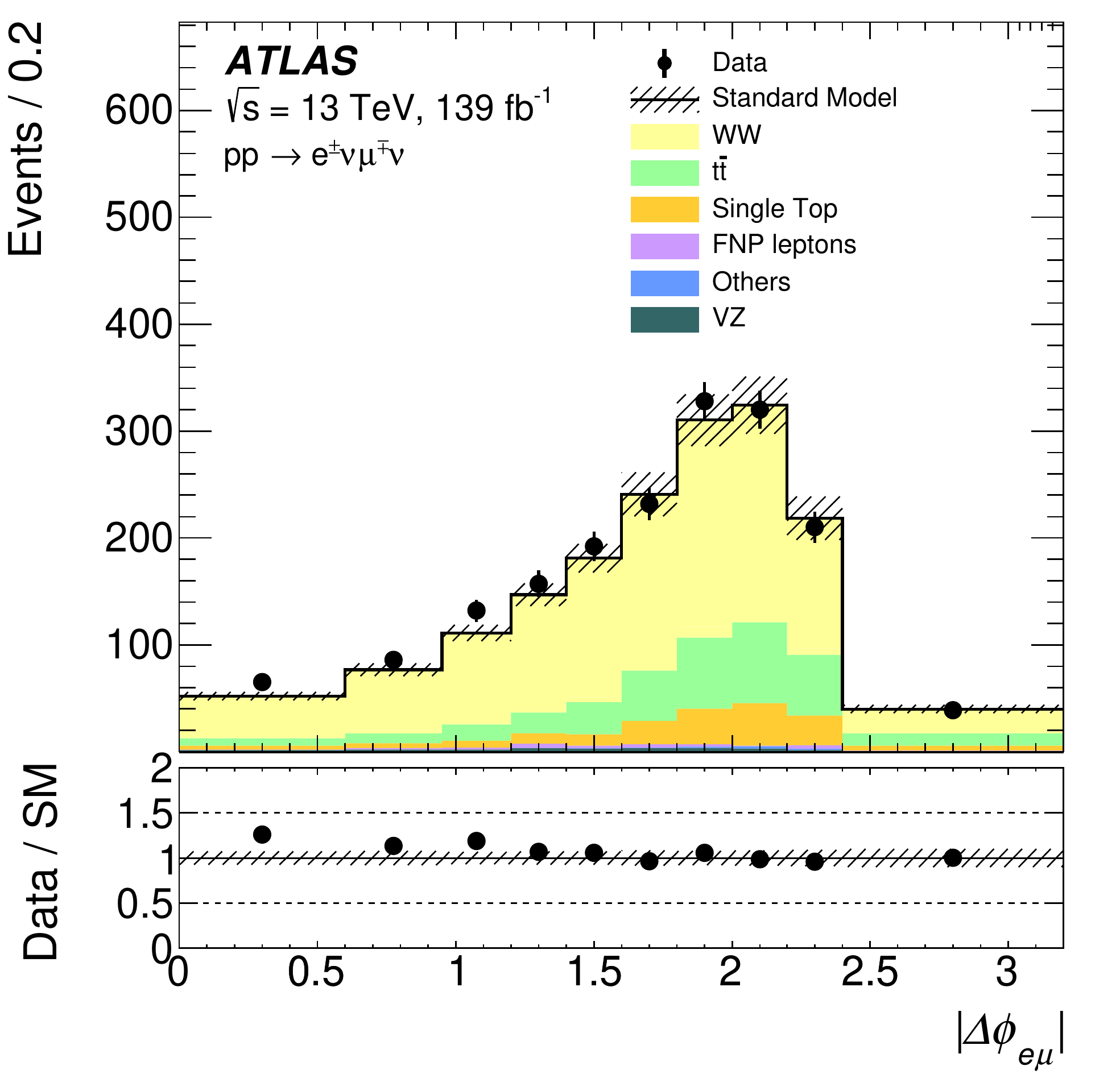} \\
\includegraphics[width=0.43\linewidth]{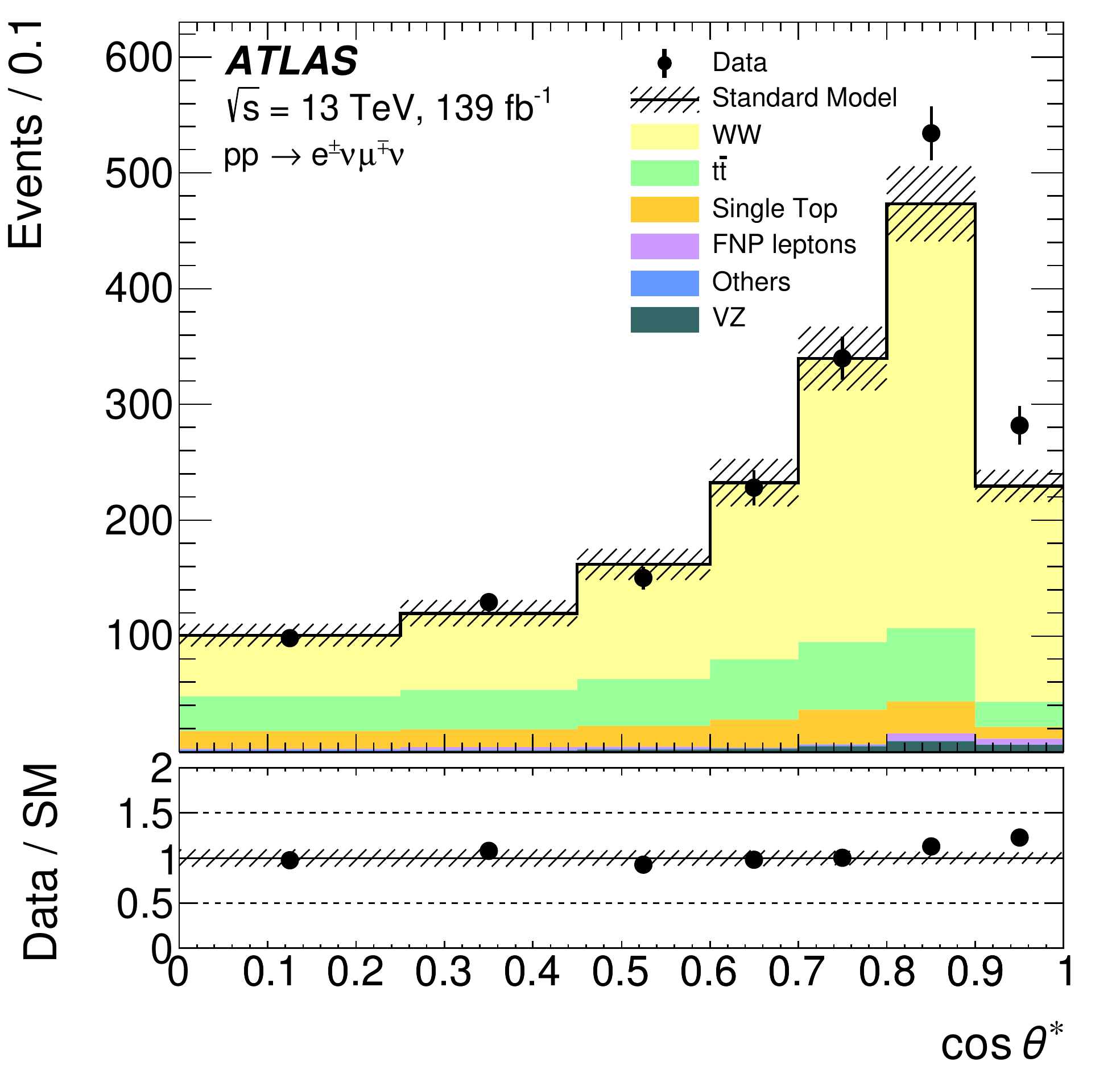}
\includegraphics[width=0.43\linewidth]{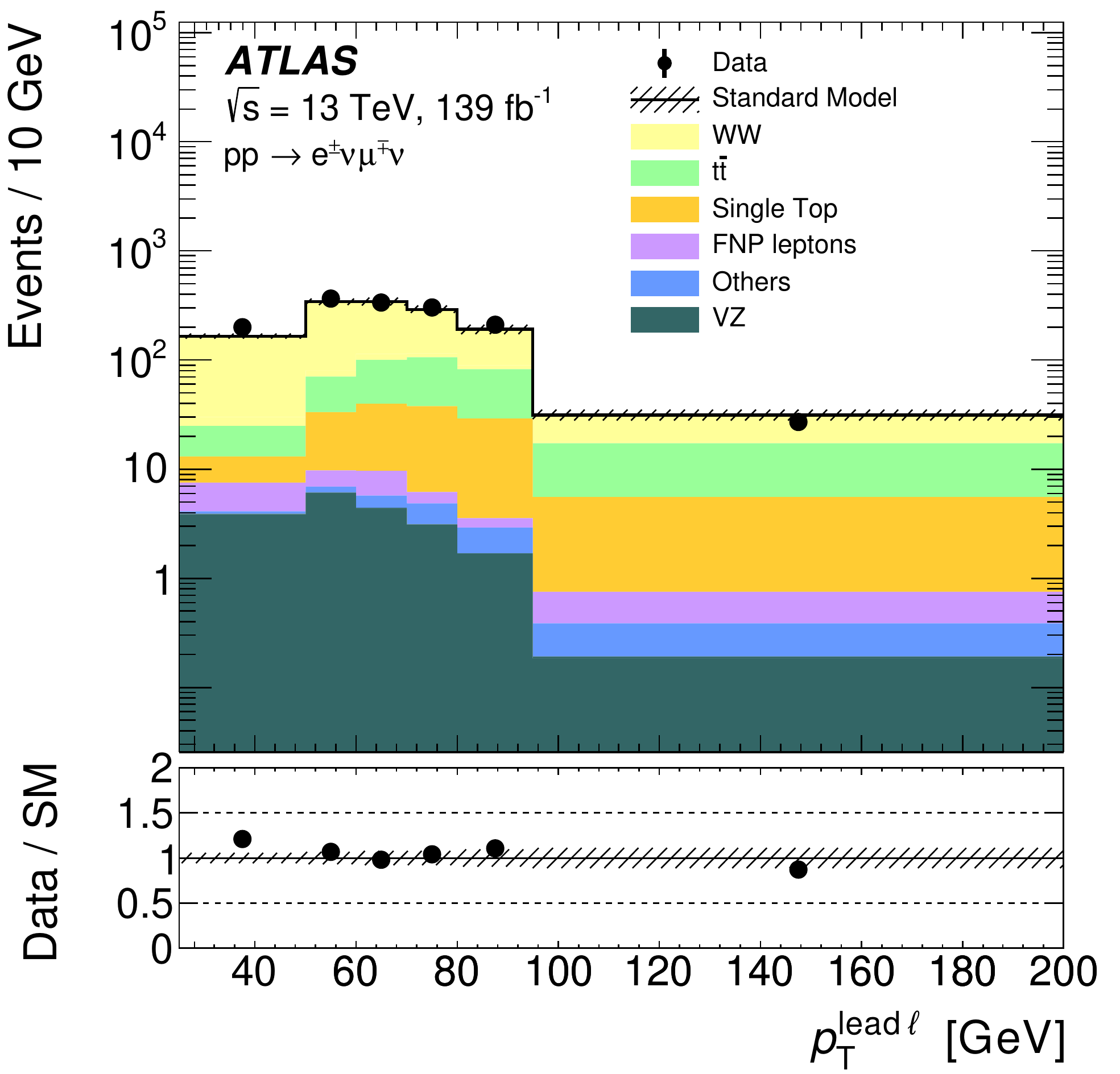} \\
\includegraphics[width=0.43\linewidth]{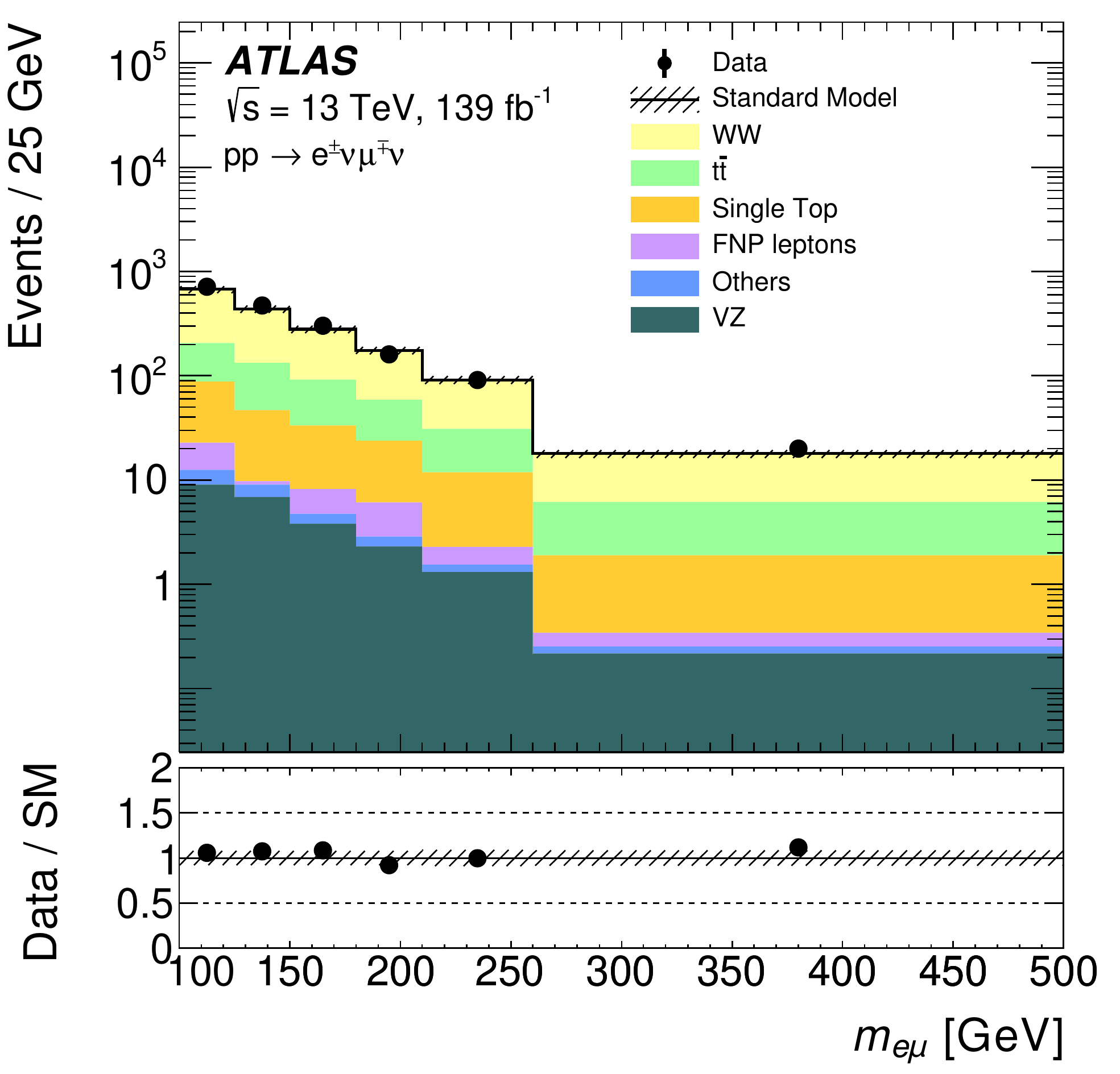}
\includegraphics[width=0.43\linewidth]{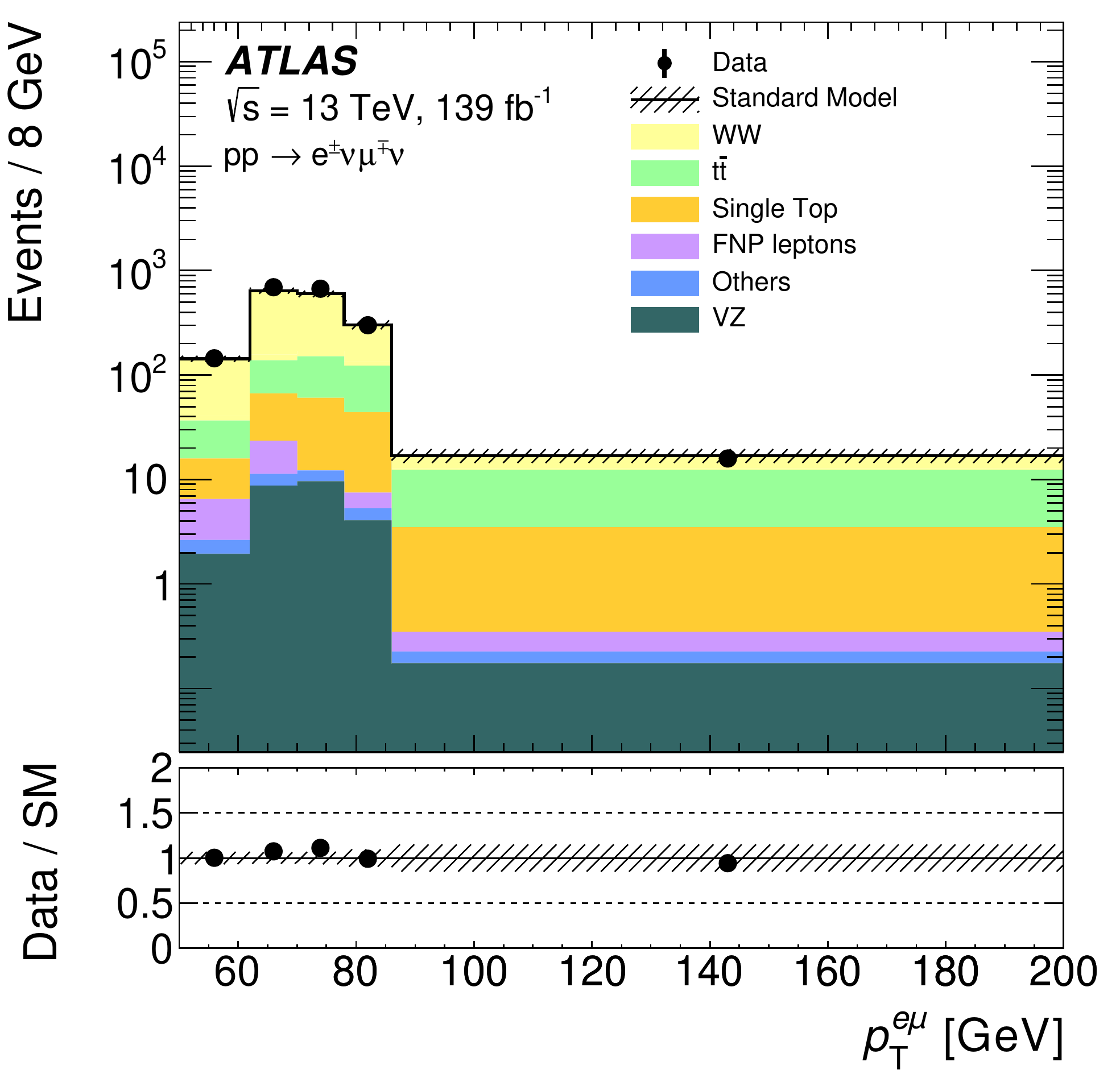} \\
 
\caption{ Signal region detector-level distributions of $|y_{e\mu}|$ (top left), $|\Delta\phi_{e\mu}|$ (top right), $\cos\theta^{\ast}$ (middle left), $p_{\mathrm{T}}^{\mathrm{lead}\, \ell}$ (middle right), $m_{e\mu}$ (bottom left), and $p_{\mathrm{T}}^{e\mu}$ (bottom right). Data are indicated by black markers along with the distribution for the $WW$ signal and background SM processes. The last bin of each scale-variable distribution contains overflow events. The lower panels show the ratio of data to the total SM background prediction. The uncertainty bands shown include statistical and systematic uncertainties, excluding theory uncertainties in the $WW$ signal. `FNP leptons' refers to the background from fake/non-prompt leptons, calculated using the data-driven matrix method.
}
\label{fig:dataMCsrww}
\end{figure}

Figure~\ref{fig:dataMCsrww} shows detector-level comparisons between the data and the SM processes for the six variables that are unfolded to particle level in this study.

\FloatBarrier
\subsection{Background estimation}
The estimation of the SM backgrounds in this study uses the same techniques as those used in the EWK 2$\ell$+0-jets search~\cite{SUSY-2018-32}. For the search, the SM backgrounds were classified into irreducible backgrounds from processes producing prompt leptons and reducible backgrounds containing one or more fake/non-prompt (FNP) leptons. The main irreducible backgrounds were SM diboson ($WW$, $WZ$, $ZZ$) and top-quark (\ttbar\ and $Wt$) production, which were estimated from simulated events and normalised using a simultaneous likelihood fit to data in dedicated control regions (CRs). The yields and shapes of kinematic distributions of the relevant backgrounds were then validated in a set of validation regions (VRs). Three CRs were used: CR-WW, targeting $WW$ production; CR-VZ, targeting $WZ$ and $ZZ$ production, which were normalised by using a single parameter in the likelihood fit to the data; and CR-top, targeting \ttbar\ and single-top-quark production,
which were also normalised by using a single parameter in the likelihood fit to the data. Both CR-VZ and CR-top require high \mttwo, and high values of \MET\ and its significance. CR-VZ uses same-flavour (dielectron and dimuon) events with a jet veto and requires the dilepton invariant mass to be consistent with an on-shell $Z$ boson. CR-top requires one electron and one muon, one $b$-tagged jet with a veto on additional non-$b$-tagged jets. The remaining background from FNP leptons was estimated from data using the matrix method~\cite{TOPQ-2010-01}. In this study, $WW$ is the target signal process, with the remaining processes being backgrounds that are subtracted from the data prior to calculating the fiducial and differential cross-sections.
 
The statistical interpretation for the search was performed using the HistFitter framework~\cite{HistFitter}. The likelihood for the `background-only' fit used to constrain the background normalisation factors was a product of Poisson probability density functions describing the observed number of events in each CR and Gaussian distributions that constrain the nuisance parameters associated with the systematic uncertainties. Poisson distributions were used for MC statistical uncertainties. Further details of the likelihood fit can be found in the EWK 2$\ell$+0-jets search paper~\cite{SUSY-2018-32}.  After the fit, the normalisation factors returned for the $WW$, $t\bar{t}$ and single-top-quark, and $WZ$/$ZZ$ processes were $1.25 \pm 0.11$, $0.82 \pm 0.06$ and $1.18 \pm 0.05$ respectively (where the errors include both statistical and systematic uncertainties), which for diboson processes were applied to MC samples scaled to NLO QCD cross-sections (the NNLO QCD cross-sections were not used in the original search paper because the samples were normalised to the data in the control regions). Good agreement, within about one standard deviation, was observed for the yields and kinematic distributions in all VRs when applying these normalisation factors and their corresponding uncertainties. The deviation of the $WW$ normalisation factor from unity by more than $1\sigma$ suggests there is tension between the SM and data in the parameter space probed by the search, and this is tested in the present study.
 
\begin{figure}[htb!]
\centering
\includegraphics[width=0.43\linewidth]{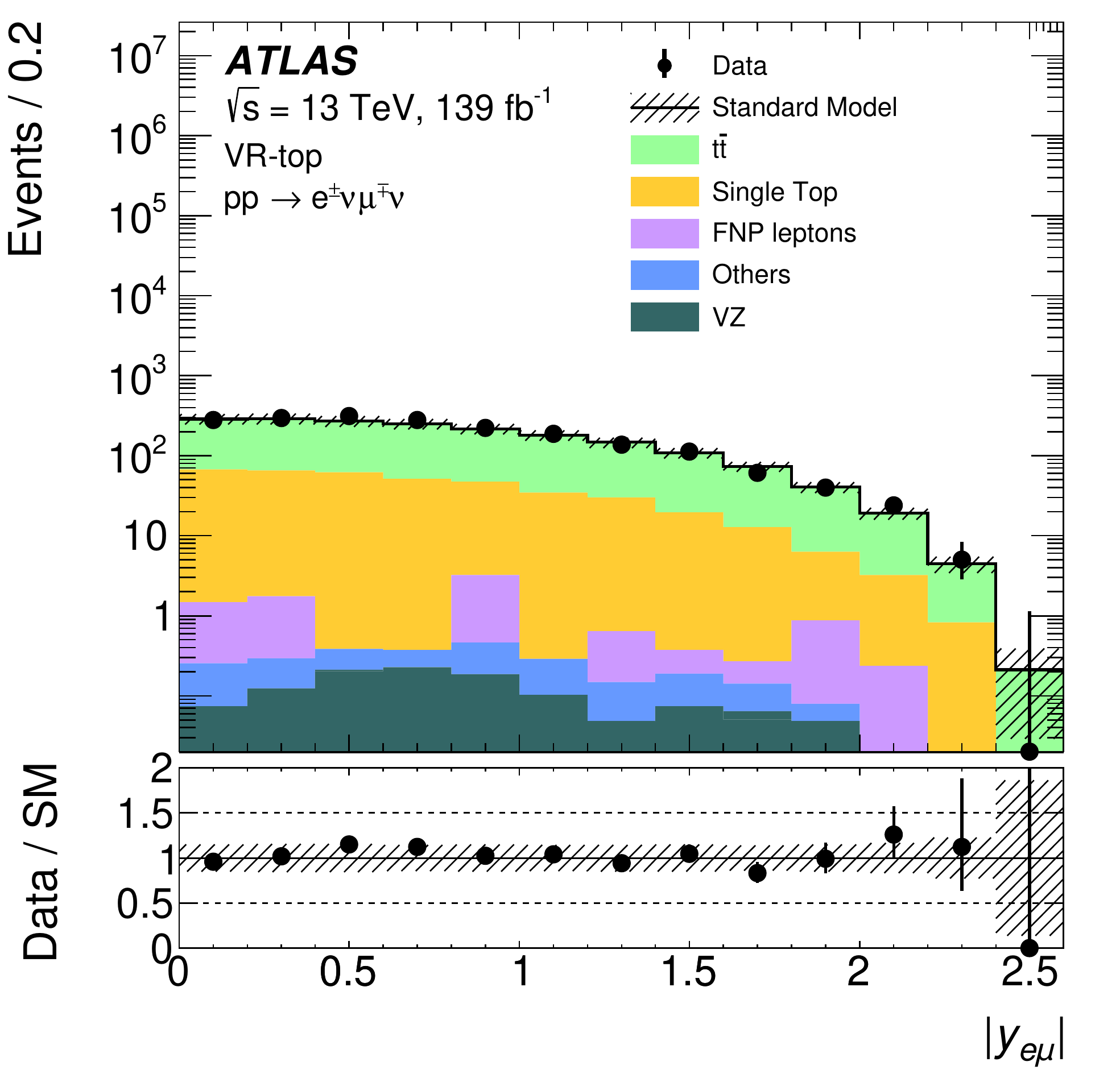}
\includegraphics[width=0.43\linewidth]{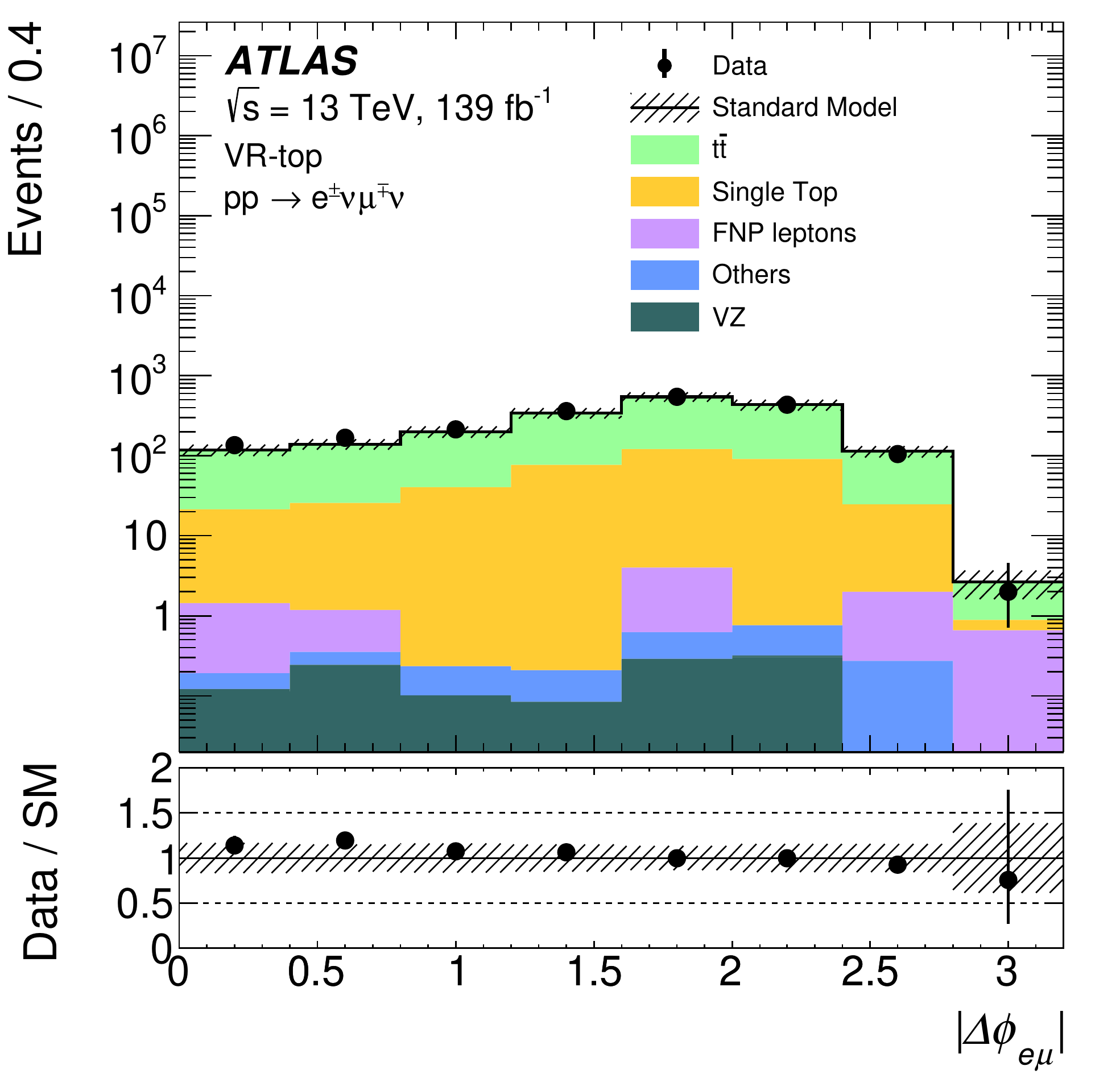} \\
\includegraphics[width=0.43\linewidth]{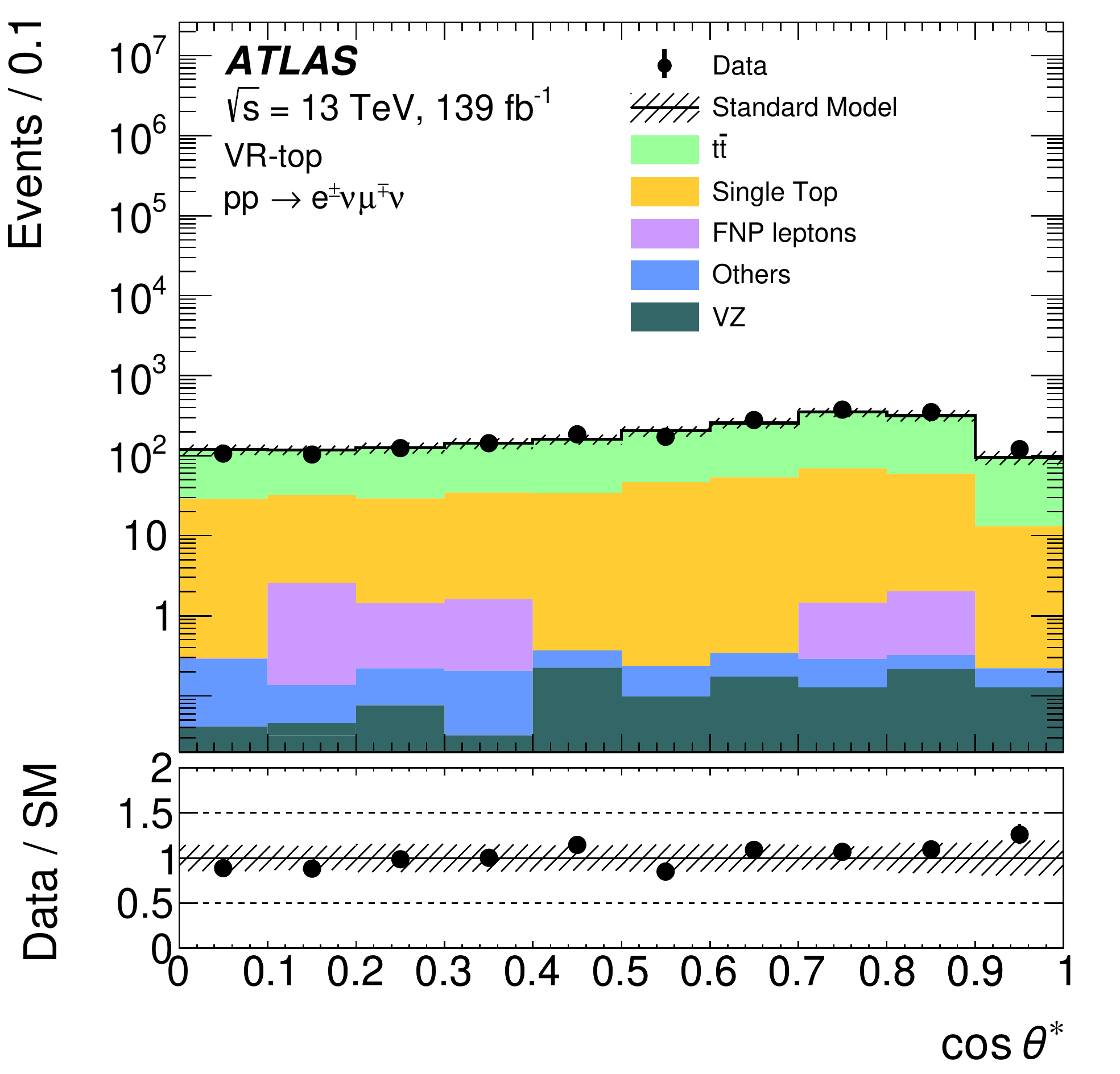}
\includegraphics[width=0.43\linewidth]{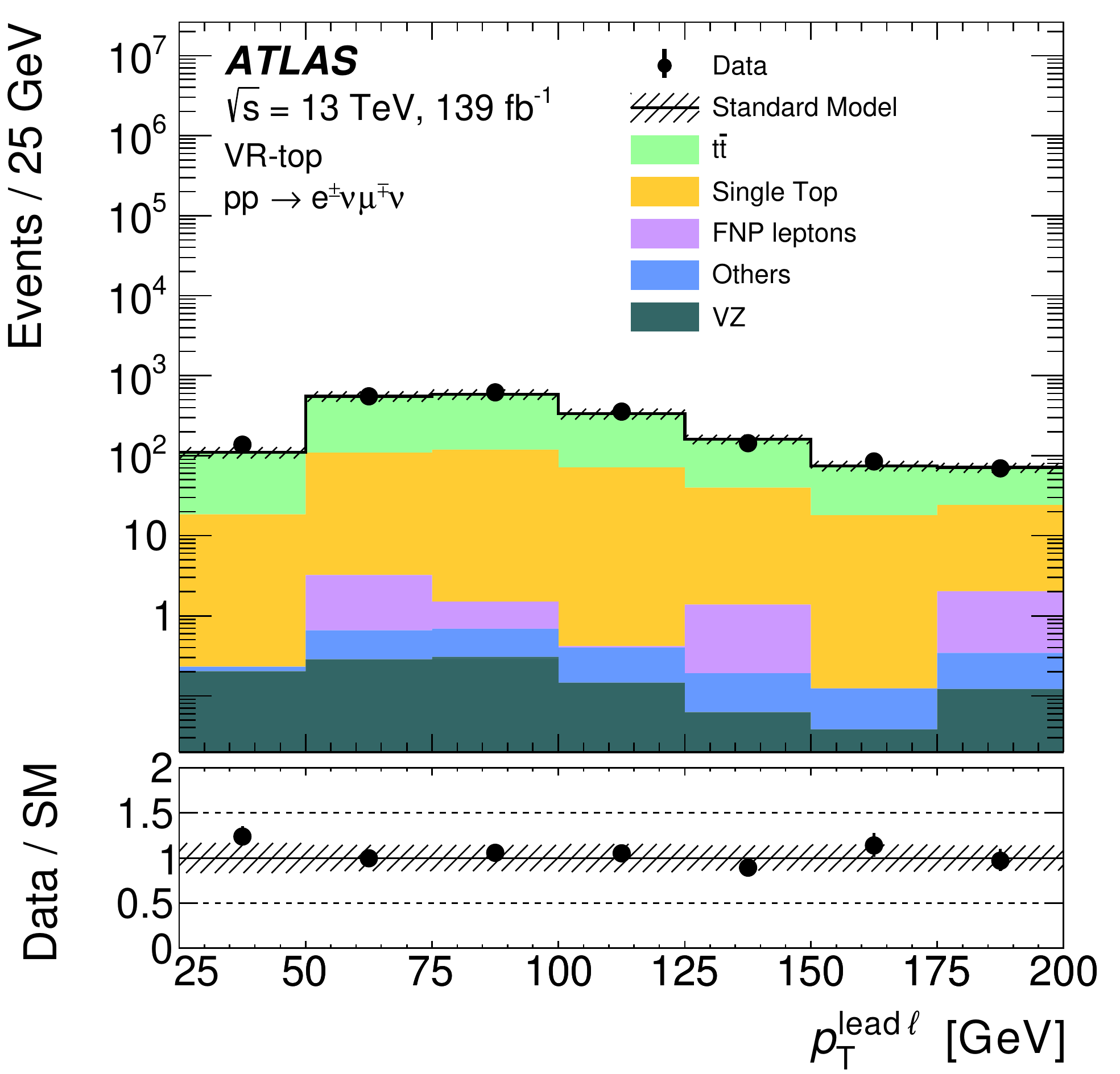} \\
\includegraphics[width=0.43\linewidth]{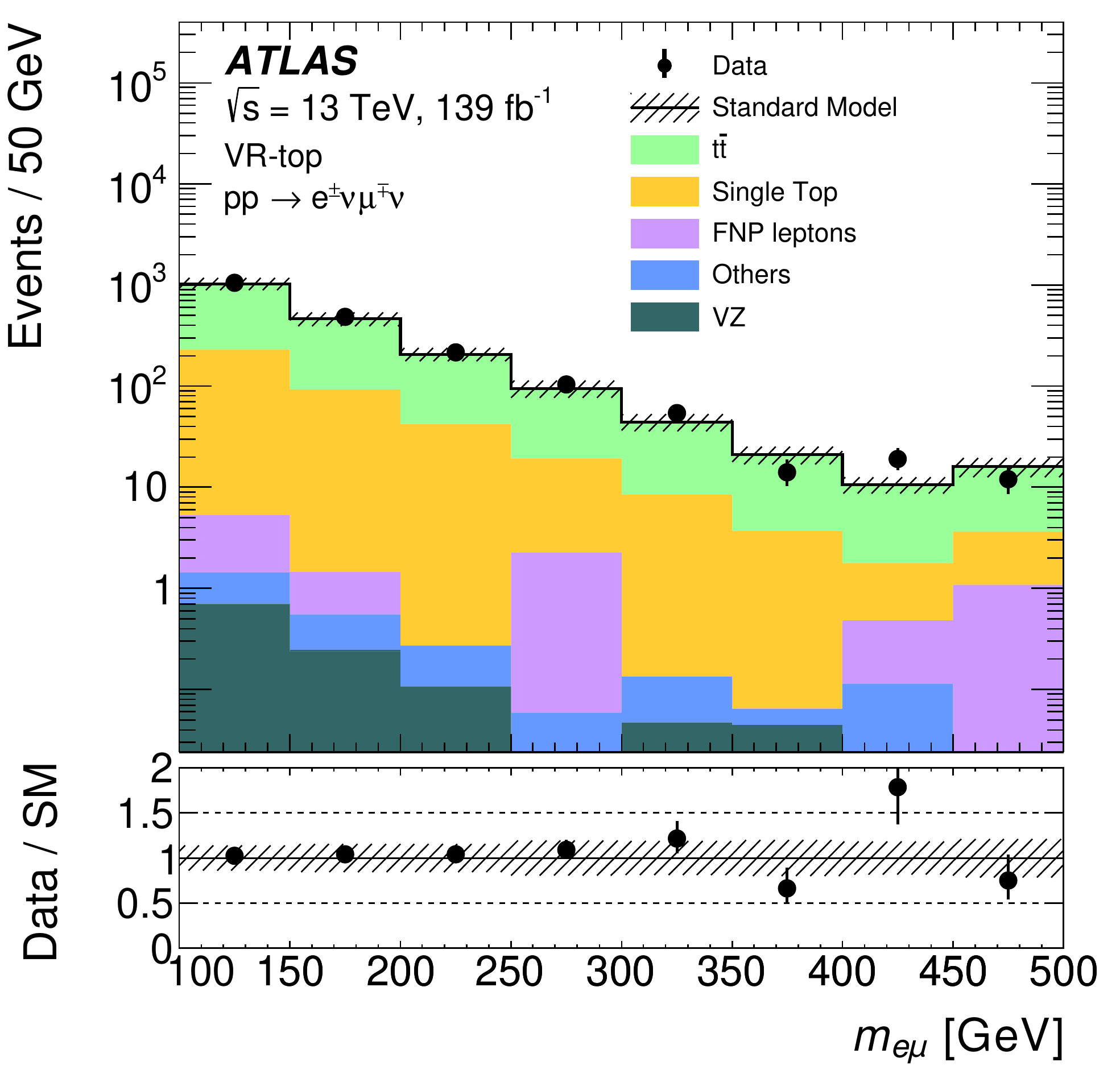}
\includegraphics[width=0.43\linewidth]{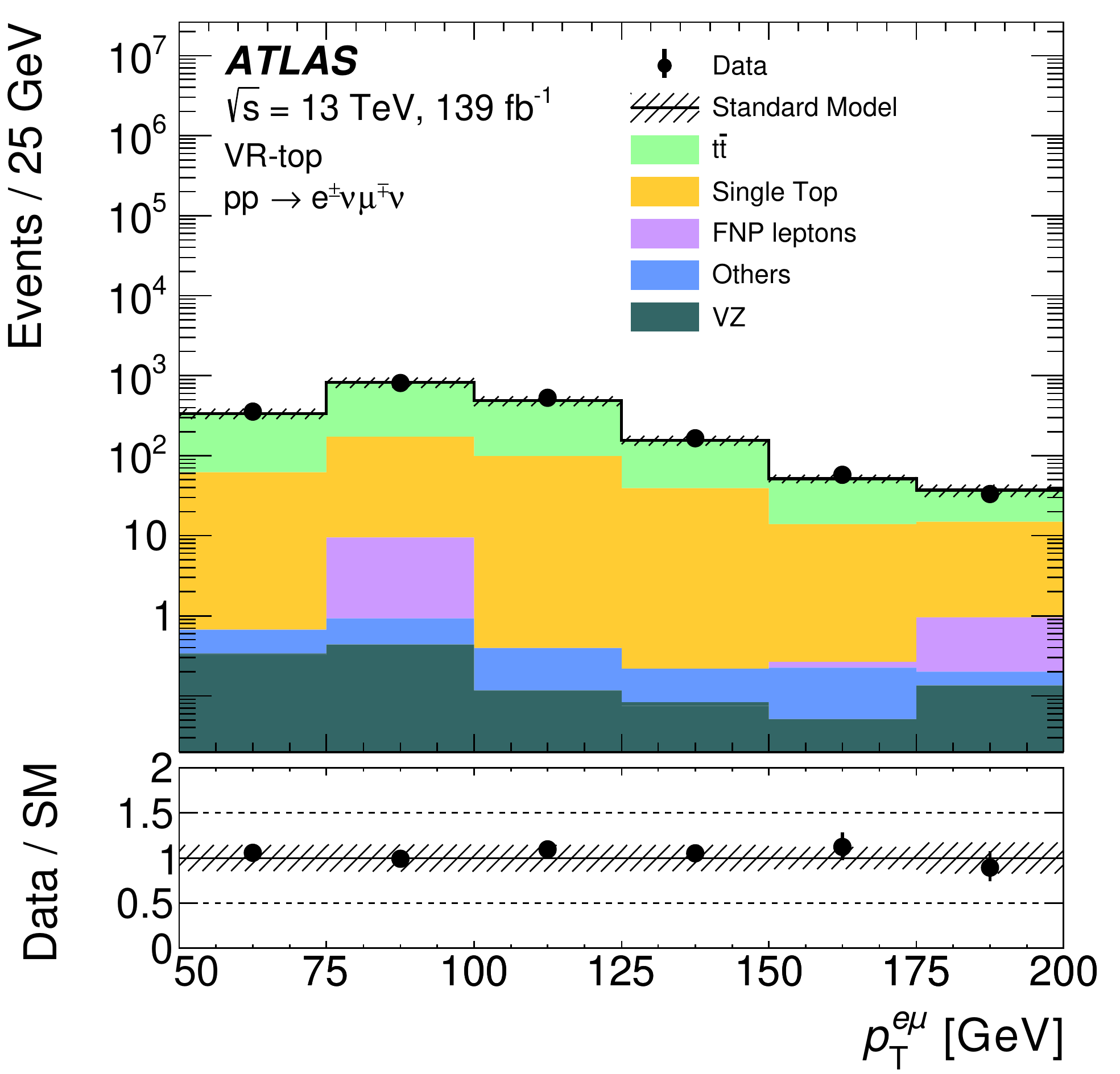} \\
 
\caption{ Detector-level distributions of $|y_{e\mu}|$ (top left), $|\Delta\phi_{e\mu}|$ (top right), $\cos\theta^{\ast}$ (middle left), $p_{\mathrm{T}}^{\mathrm{lead}\, \ell}$ (middle right), $m_{e\mu}$ (bottom left), and $p_{\mathrm{T}}^{e\mu}$ (bottom right) in the top validation region. Data are indicated by black markers along with the distribution for the $WW$ signal and background SM processes. The last bin of each scale-variable distribution contains overflow events. The lower panels show the ratio of data to the total SM background prediction. The uncertainty bands shown include statistical and systematic uncertainties, excluding theory uncertainties in the $WW$ signal. `FNP leptons' refers to the background from fake/non-prompt leptons, calculated using the data-driven matrix method.
}
\label{fig:topvrplots}
\end{figure}

In this study, the normalisation factors from the EWK 2$\ell$+0-jets search are applied directly to the $VZ$ ($WZ$/$ZZ$) and top ($t\bar{t}$,$Wt$) backgrounds that are subtracted from the data when performing the cross-section measurements described in Section~\ref{sec:unfolding}. The uncertainties in the normalisation factors are propagated through the calculation as discussed in Section~\ref{sec:systematics}. The correlations and constraints that the likelihood fit imposes on the nuisance parameters describing the systematic uncertainties are not applied in this study; the systematic uncertainties are instead assumed to take their nominal values as discussed in Section~\ref{sec:systematics}. This approach is designed to be conservative, but it has negligible impact on the results because no significant constraints were observed in the EWK 2$\ell$+0-jets search. To validate the use of the original top normalisation factor from the EWK 2$\ell$+0-jets search in the adjusted phase space of this study, an additional validation exercise is performed to check the modelling of the top-quark background in a region with the same selection as in Table~\ref{tab:unfoldingregions} but requiring exactly one $b$-tagged jet. Good agreement is observed across all six distributions considered for differential cross-section measurements, as shown in Figure~\ref{fig:topvrplots}.

\FloatBarrier

\subsection{Fiducial cross-section determination}
\label{sec:unfolding}

The differential cross-sections are measured in the fiducial phase space of the $WW \rightarrow e^{\pm}\nu\mu^{\mp}\nu$ decay channel using particle-level implementations of the selection criteria defined in Table~\ref{tab:unfoldingregions}. The particle-level quantities associated with simulated events are calculated using the SimpleAnalysis~\cite{ATL-PHYS-PUB-2022-017} framework. The signal particle-level distributions produced by SimpleAnalysis have been validated against the Rivet~\cite{Bierlich:2019rhm} toolkit that enables further reinterpretation of SM measurements and validation of MC generators. The Rivet routine for this measurement is available on HepData~\cite{Maguire:2017ypu}.
 
Electrons and muons are required to originate from the hard interaction and not from hadron decays.  Electrons and muons from leptonically decaying $\tau$-leptons are included in the fiducial region. The momenta of photons that are emitted in a cone of size $\Delta R'=0.1$ around the lepton direction and do not originate from hadron decays are added to form `dressed' leptons. Particle-level jets are reconstructed using the anti-$k_t$ algorithm~\cite{Cacciari:2008} with radius parameter $R=0.4$ from visible stable final-state particles, excluding prompt dressed leptons. The particle-level missing transverse momentum is defined as
the vectorial sum of the momenta of invisible particles in the event. For SM processes this is the sum of the neutrino momenta.
 
The fiducial cross-section is calculated as:
\begin{displaymath}
\sigma_{WW} =  \frac{N_{\text{obs}}-N_{\text{bkg}}}{C\cdot\mathcal{L}},
\end{displaymath}
where $N_{\text{obs}}$ is the observed number of data events in the fiducial region, $N_{\text{bkg}}$ is the predicted number of background events,  $\mathcal{L}$ the integrated luminosity, and $C$ is a correction factor to account for limited acceptances and detector inefficiencies. It is calculated using MC simulation as the number of simulated signal events passing the detector-level event selection divided by the number of events in the fiducial phase space. In this study, $C=0.55 \pm 0.08$ is applied, where the uncertainty comes from statistical, experimental and theoretical sources, as described in Section~\ref{sec:systematics}.
 
The differential cross-sections are calculated using the iterative Bayesian unfolding  (IBU) technique~\cite{DAgostini:1995,BayesUnfold2} as implemented in the RooUnfold package~\cite{Adye:2011gm}. This unfolding technique corrects the detector-level distributions of data (with the non-$WW$ backgrounds subtracted) for bin-to-bin migrations due to the event reconstruction. It also applies fiducial corrections (corresponding to events that are reconstructed in the signal region but originate outside the fiducial region at particle level) and reconstruction efficiency corrections (due to events that lie inside the fiducial region at particle level but do not enter the signal region due to detector inefficiencies). The bins chosen for the differential measurements were optimised for a desired level of statistical uncertainty and to reduce the migration of events between particle-level and detector-level bins. The number of iterations used in IBU is also optimised by considering the bias due to the assumed true distribution and the resulting statistical uncertainty of the measurement, with too many iterations generating high statistical uncertainties and too few iterations biasing the measurements towards the MC prediction. In this study, two iterations are chosen for $\cos\theta^{\ast}$ and $|\Delta\phi_{e\mu}|$, three iterations are used for $m_{e\mu}$ and $p_{\mathrm{T}}^{\mathrm{lead}\, \ell}$, and four iterations are used for $|y_{e\mu}|$ and $p_{\mathrm{T}}^{e\mu}$. In addition to the bias tests (discussed in Section~\ref{sec:systematics}) to measure any systematic effects due to the use of the signal $WW$ MC sample in the unfolding procedure, several signal injection tests were performed using SUSY models for chargino-pair production that were on the edge of the exclusion sensitivity in the EWK 2$\ell$+0-jets search. These are important checks of the validity of using these measurements to calculate constraints on BSM physics. Detector-level distributions of $WW$ plus injected BSM signal were input to the unfolding calculation to test whether the particle-level $WW$ plus BSM distribution could be recovered. The unfolding calculation matched the expected $WW$ plus BSM distributions for a range of SUSY models displaying different kinematics because of their different SUSY particle masses. The results of the BSM injection tests are available on HepData~\cite{Maguire:2017ypu}.

\subsection{Systematic uncertainties}
\label{sec:systematics}
Systematic uncertainties in the $WW$ differential cross-sections measured in this study arise from experimental sources (which impact the subtracted non-$WW$ backgrounds, and the calculation used to correct the signal for detector effects), uncertainties in the modelling of the top-quark background (which includes theoretical uncertainties, and uncertainties associated with the data-driven background estimate), and signal modelling. Statistical uncertainties associated with the MC samples used for the signal and background processes, and with the observed data distributions, also impact the unfolded distributions.
 
The sources of experimental uncertainty considered in the EWK 2$\ell$+0-jets search~\cite{SUSY-2018-32} are also considered in this study. The dominant experimental uncertainties are due to the calibration of the jet energy scale and resolution~\cite{PERF-2016-04,JETM-2018-05}. Additional uncertainties that arise from the lepton reconstruction efficiency, lepton energy scale and lepton energy resolution, and differences between the trigger efficiencies in data and simulation are grouped into the lepton uncertainties category. There are also uncertainties in the scale factors applied to the simulated samples to account for differences between data and simulation in the $b$-jet identification efficiency, and an uncertainty in $\mathbf p^\mathrm{miss}_\mathrm{T}$ associated with the soft-term resolution and scale
~\cite{PERF-2016-07}. Finally, an uncertainty is assigned to the reweighting procedure (pile-up reweighting) applied to simulated events to match the distribution of the number of
interactions per bunch crossing observed in data.
 
Several sources of uncertainty in the modelling of $t\bar{t}$ and $Wt$ events are accounted for by varying the normalisation and shape of the subtracted backgrounds. For \ttbar production, uncertainties in the parton shower simulation are estimated from differences between samples generated with \POWHEGBOX interfaced to either \PYTHIA[8.186] or \HERWIG[7.04]~\cite{Bahr:2008pv,Bellm:2015jjp}. Uncertainties in the modelling of initial- and final-state radiation are estimated by comparing the nominal sample with two alternative samples generated with \POWHEGBOX interfaced to \PYTHIA[8.186] but with the radiation settings varied~\cite{ATL-PHYS-PUB-2016-020}. Finally, an additional uncertainty associated with the choice of event generator is estimated by comparing the nominal samples with samples generated with \MGNLO\ interfaced to \PYTHIA[8.186]~\cite{ATL-PHYS-PUB-2017-007}. For single-top-quark production, an uncertainty is assigned to the treatment of the interference between the $Wt$ and $t\bar{t}$ samples. This is done by comparing the nominal sample generated using the diagram removal method with a sample generated using the diagram subtraction method~\cite{ATL-PHYS-PUB-2016-020}.
 
Of the systematic uncertainties considered in the EWK 2$\ell$+0-jets search, uncertainties in the data-driven estimate of FNP leptons and theoretical uncertainties in the diboson $WZ/ZZ$ backgrounds are not applied in this study because these processes contribute little to the subtracted backgrounds. 
Additional systematic uncertainties are applied to the unfolding to account for the uncertainty in the normalisation of the top-quark and $VZ$ backgrounds, although the $VZ$ normalisation uncertainties are observed to be negligible. The luminosity uncertainty (1.7\%) is applied to the subtracted backgrounds that are not estimated using data-driven techniques.
 
Tests were performed to estimate the bias introduced by using information from the nominal signal MC sample in the unfolding procedure. This includes a data-driven test, whereby MC simulated $WW$ signal events are reweighted at generator level to obtain better agreement between the detector-level signal and the background-subtracted data. The nominal unfolding procedure is then applied to the reweighted detector-level signal distributions to check whether the reweighted particle-level distributions can be reproduced. The impact of theoretical uncertainties in the signal modelling is evaluated by using the detector-level signal distributions with the alternative \SHERPA $qq\rightarrow WW$ signal sample introduced in Section~\ref{sec:samples} as input to the nominal unfolding procedure, and comparing the result with the alternative particle-level signal distribution. In all tests the expected particle-level distributions were accurately recovered so no additional uncertainties were assigned to the unfolding procedure.
 
Finally, statistical uncertainties from the data are calculated using pseudo-experiments that vary the data distributions according to their Poisson uncertainties in each bin, which are then passed through the unfolding calculation. Statistical uncertainties associated with the simulated MC samples are evaluated using a similar technique.
 
\section{Results}
\label{sec:results}
 
The measured fiducial cross-section for $WW \rightarrow e^{\pm}\nu\mu^{\mp}\nu$ production in the phase space defined in Table~\ref{tab:unfoldingregions} is:
\begin{displaymath}
\sigma(WW\rightarrow e^{\pm}\nu\mu^{\mp}\nu) = 19.2 \pm 0.3~(\mathrm{stat}) \pm 2.5~(\mathrm{syst}) \pm 0.4~(\mathrm{lumi})~\mathrm{fb} = 19.2 \pm 2.6~(\mathrm{total})~\mathrm{fb}.
\end{displaymath}
Table ~\ref{tab:total_xsec_systs} shows the relative impact of the categories of systematic uncertainties discussed in Section~\ref{sec:systematics} on the measured fiducial cross-section. The largest contribution is from the experimental jet uncertainty, which contributes a $12\%$ uncertainty to the measured fiducial cross-section. The jet uncertainties are higher than in the previous ATLAS 13~\TeV\ $WW$+0-jet measurement~\cite{STDM-2017-24} and this can be attributed to the lower \pT\ threshold used to define the jet veto. 
 
\begin{table}[htb!]
\caption{\label{tab:total_xsec_systs} Breakdown of the relative uncertainties per category and the total uncertainty on the fiducial cross-section measurement.}
\small
\centering
\begin{tabular}{lc}
\hline
Uncertainty source & Uncertainty [\%] \\
\hline
Jets & 11.7 \hspace{3pt} \\
Top modelling & 4.8 \\
Data statistics & 3.1 \\
Lepton modelling & 1.9 \\
Luminosity & 1.7 \\
Pile-up reweighting & 1.2 \\
\MET modelling & 1.1 \\
MC statistical uncertainties & 0.5 \\
\hline
Total systematic uncertainty & 13.0  \hspace{3pt} \\
\hline
Total uncertainty & 13.4  \hspace{3pt}  \\
\hline
\end{tabular}
\end{table}
 
The measured value is compatible with the nominal predictions of 17.8~fb and 17.1~fb from \POWHEGBOX[v2] +\,\PYTHIA[8.186] and  \SHERPA[2.2.2], respectively, where both are combined with \SHERPA[2.2.2] +\,\OPENLOOPS (LO+PS) for the $gg$-initiated states. The ratio of the measured cross-section to the nominal \POWHEGBOX[v2] +\,\PYTHIA[8.186] prediction is 1.08. 
To compare this ratio with the detector-level $WW$ normalisation factor of $1.25 \pm 0.11$ in the EWK 2$\ell$+0-jets search~\cite{SUSY-2018-32}, the former must be multiplied by 1.13 to account for the NLO cross-section calculation, which is included in this study but not in the EWK 2$\ell$+0-jets search. This gives a ratio of 1.22, which is consistent with the normalisation factor from the EWK 2$\ell$+0-jets search.
 
\begin{figure}[htb!]
\centering
\includegraphics[width=0.41\linewidth]{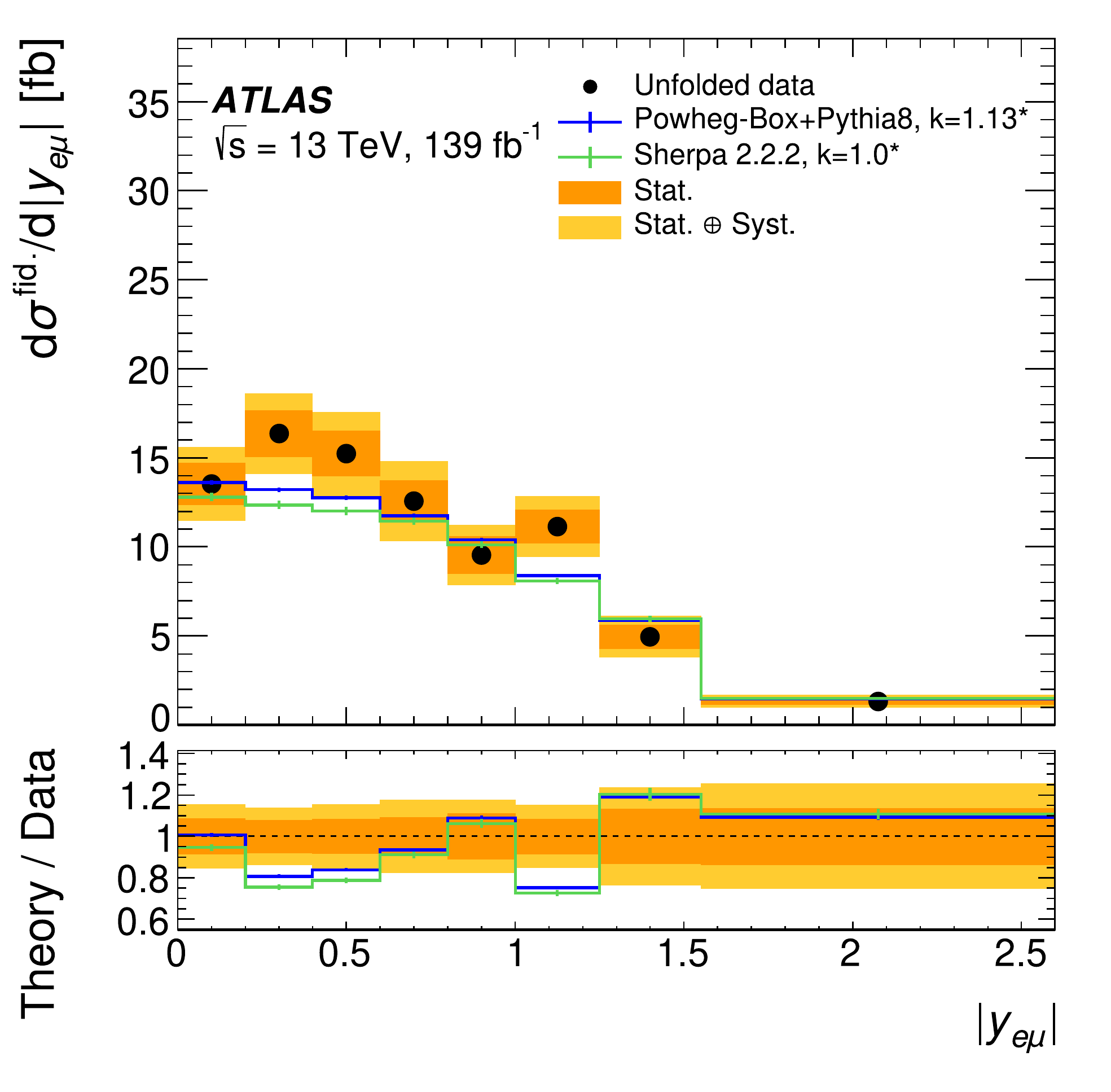}
\includegraphics[width=0.41\linewidth]{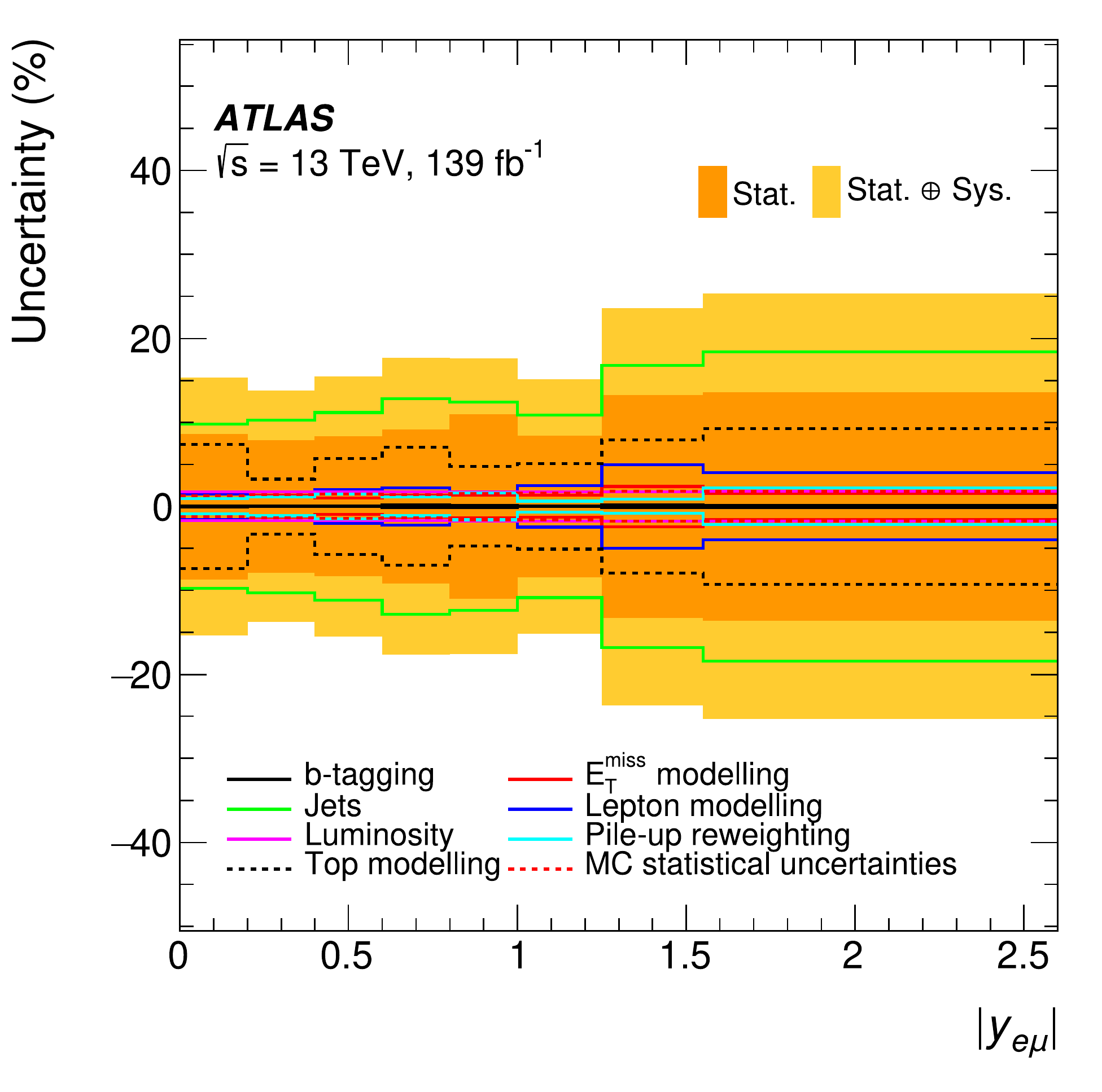} \\
 
\includegraphics[width=0.41\linewidth]{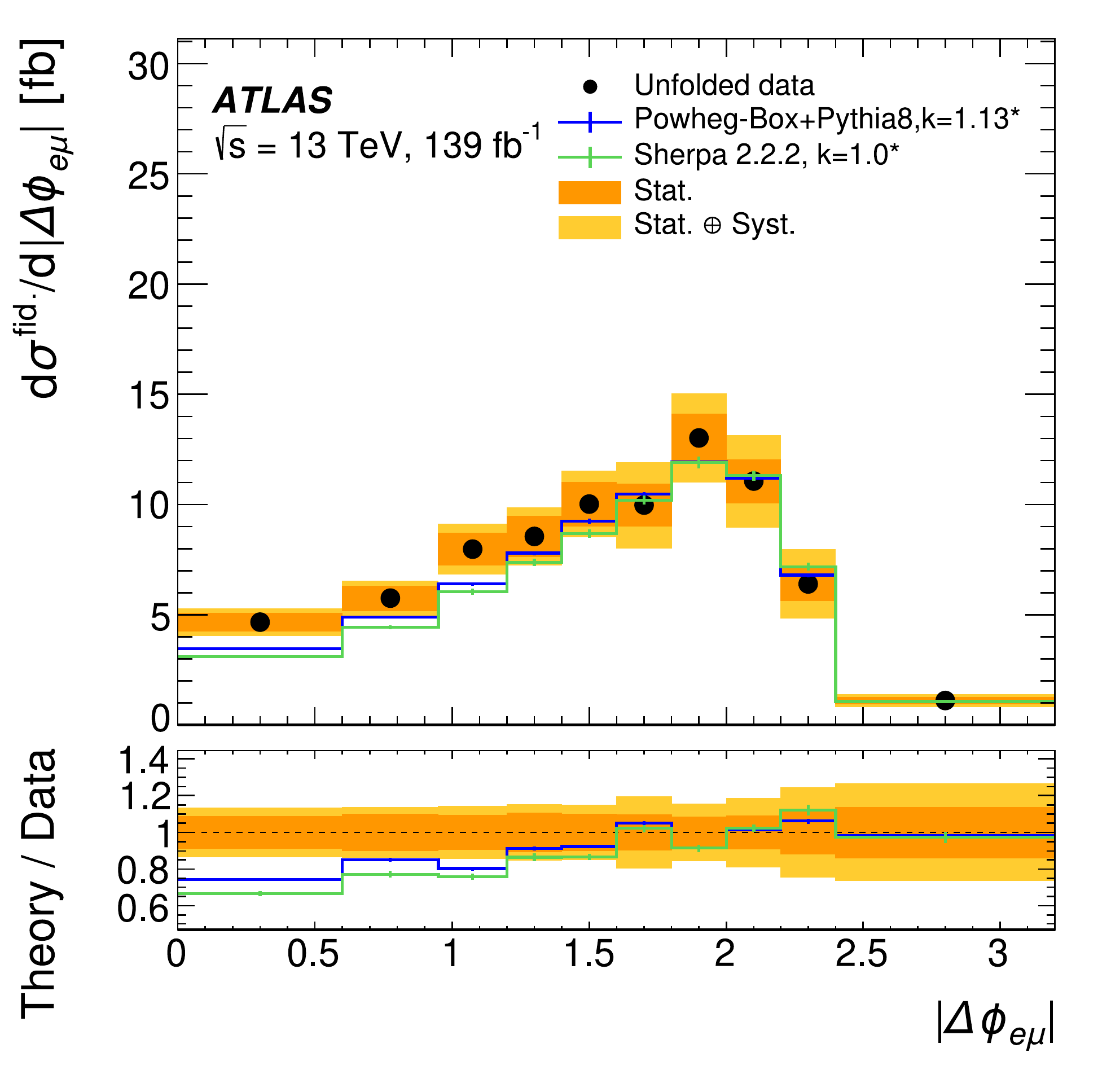}
\includegraphics[width=0.41\linewidth]{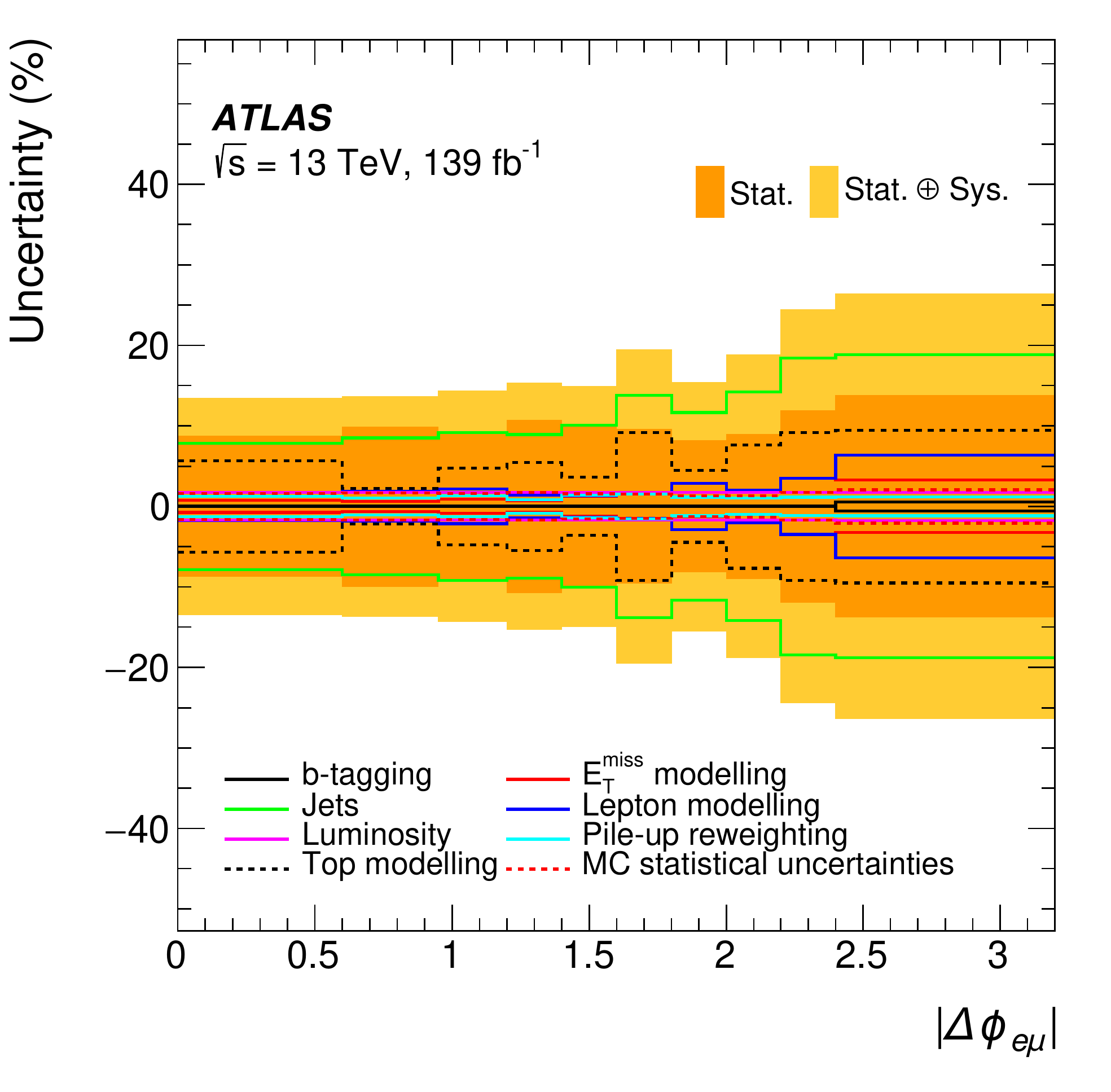} \\
 
\includegraphics[width=0.41\linewidth]{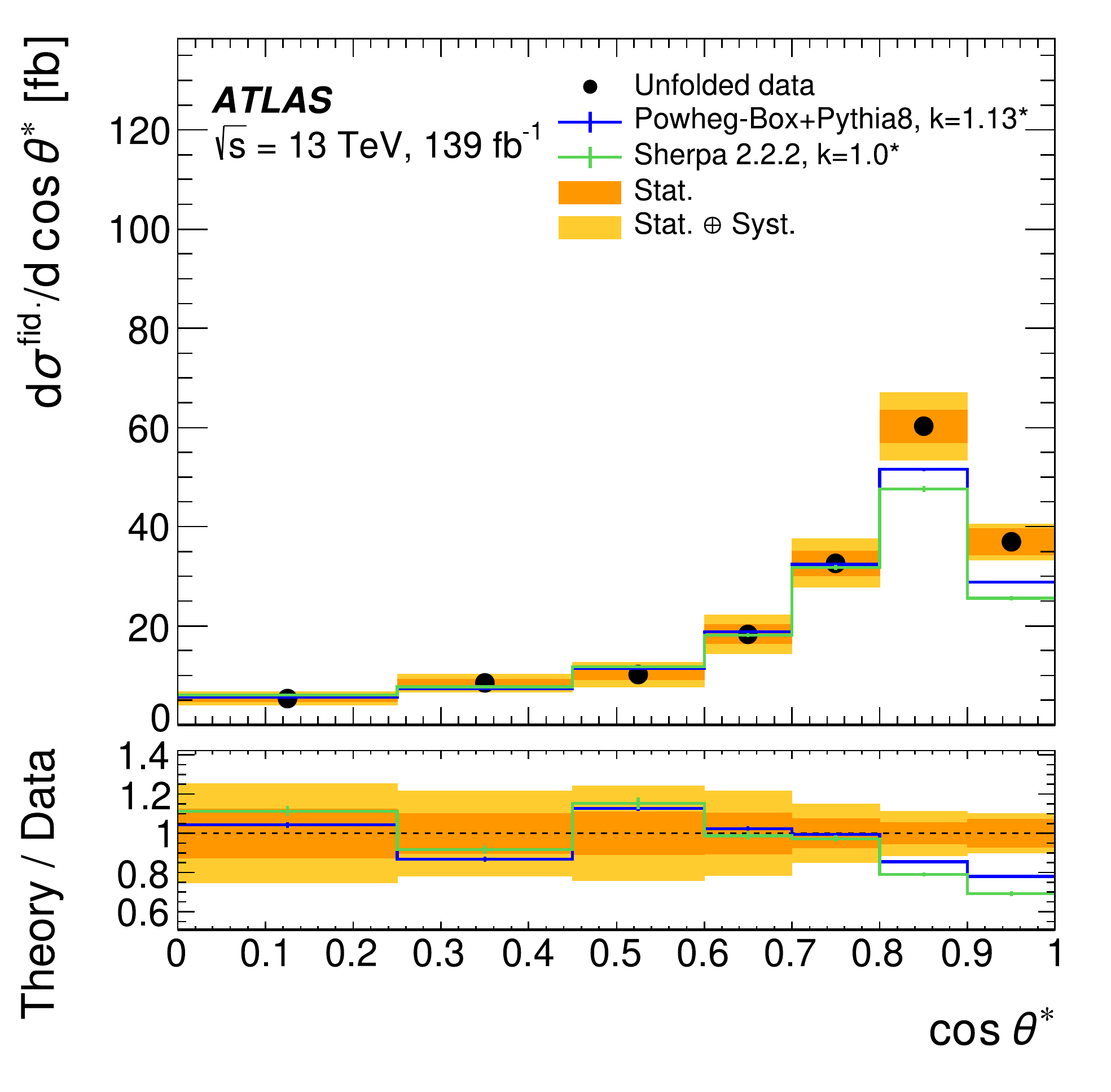}
\includegraphics[width=0.41\linewidth]{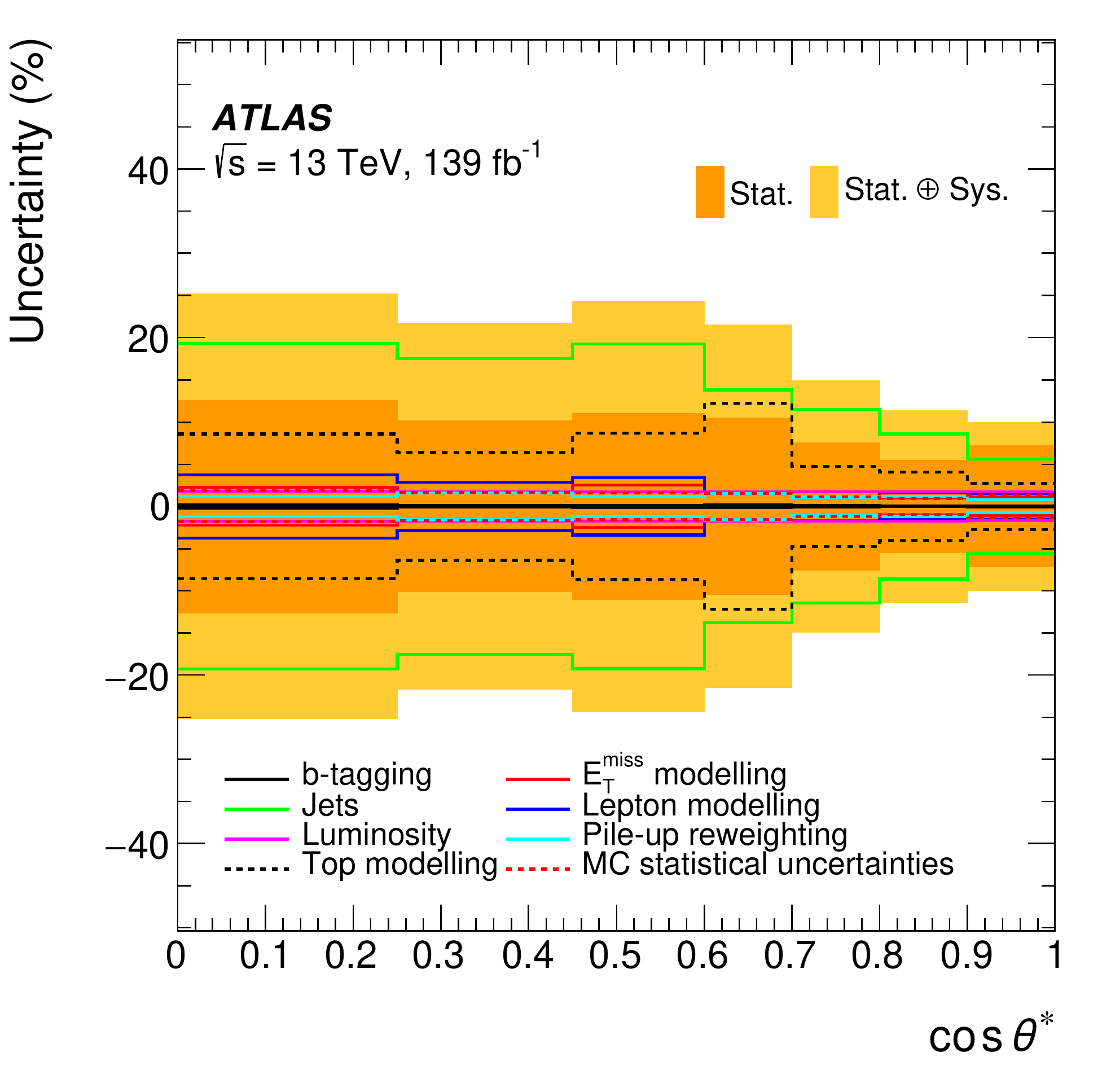} \\
 
\caption{Measured fiducial differential cross-sections of $WW$ production for (top to bottom)  $|y_{e\mu}|$, $|\Delta\phi(e\mu)|$ and $\cos\theta^{\ast}$. The measured cross-section values are shown as points with dark bands giving the statistical uncertainty and light bands indicating the size of the total uncertainty. The results are compared with the $q\bar{q}$-initiated predictions from \POWHEGBOX[v2] +\,\PYTHIA[8.186] and \SHERPA[2.2.2], each combined with  \SHERPA[2.2.2] +\,\OPENLOOPS (LO+PS) for the $gg$-initiated states. The $k$-factors refer to the corrections applied to scale the predictions of $qq$-initiated and $gg$-initiated processes to NNLO and NLO accuracy in QCD respectively. The right column shows a breakdown of contributions to the uncertainties in the unfolded measurement. }
\label{fig:unfoldedresultsAng}
\end{figure}

\begin{figure}[htb!]
\centering
\includegraphics[width=0.41\linewidth]{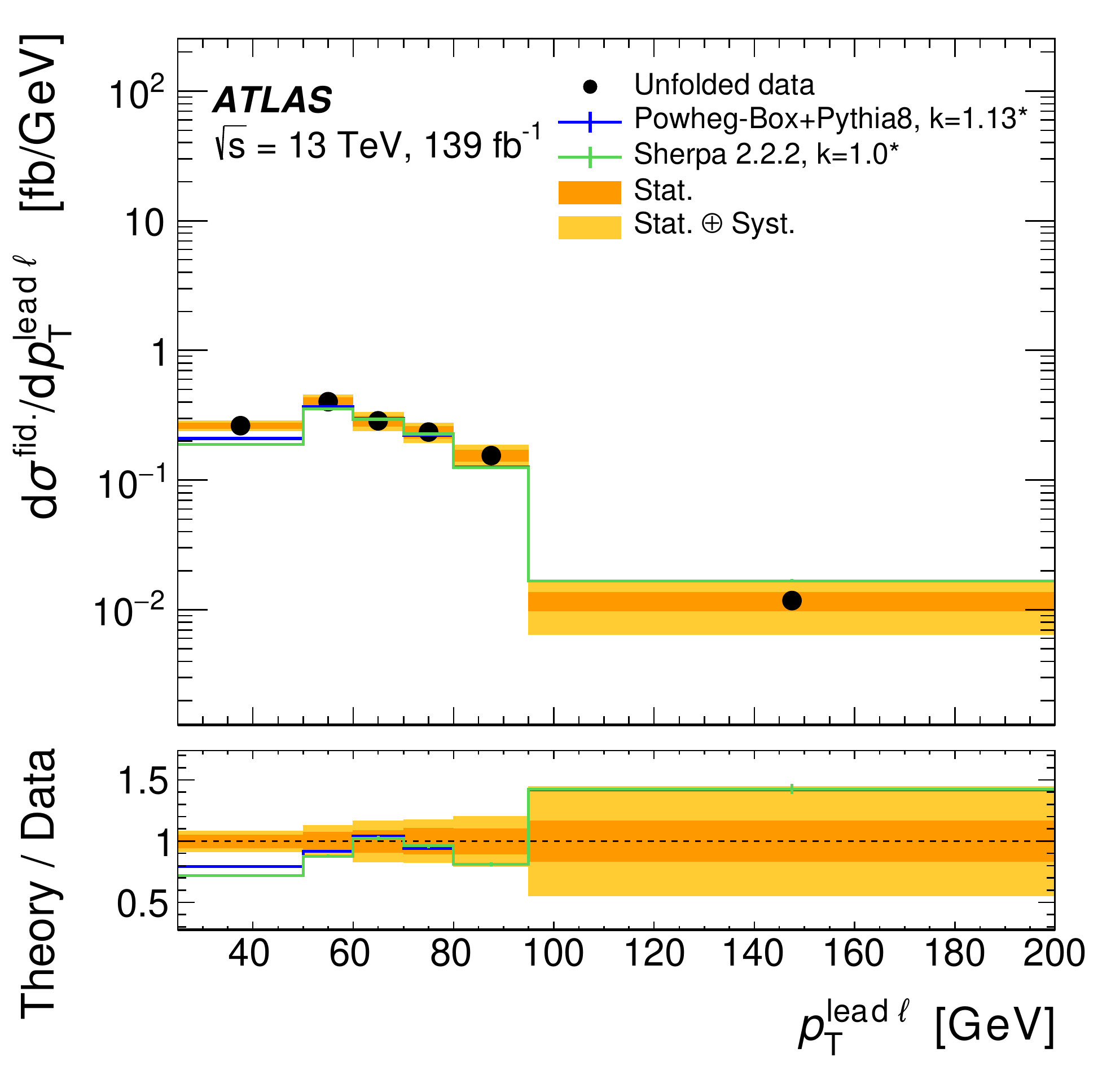}
\includegraphics[width=0.41\linewidth]{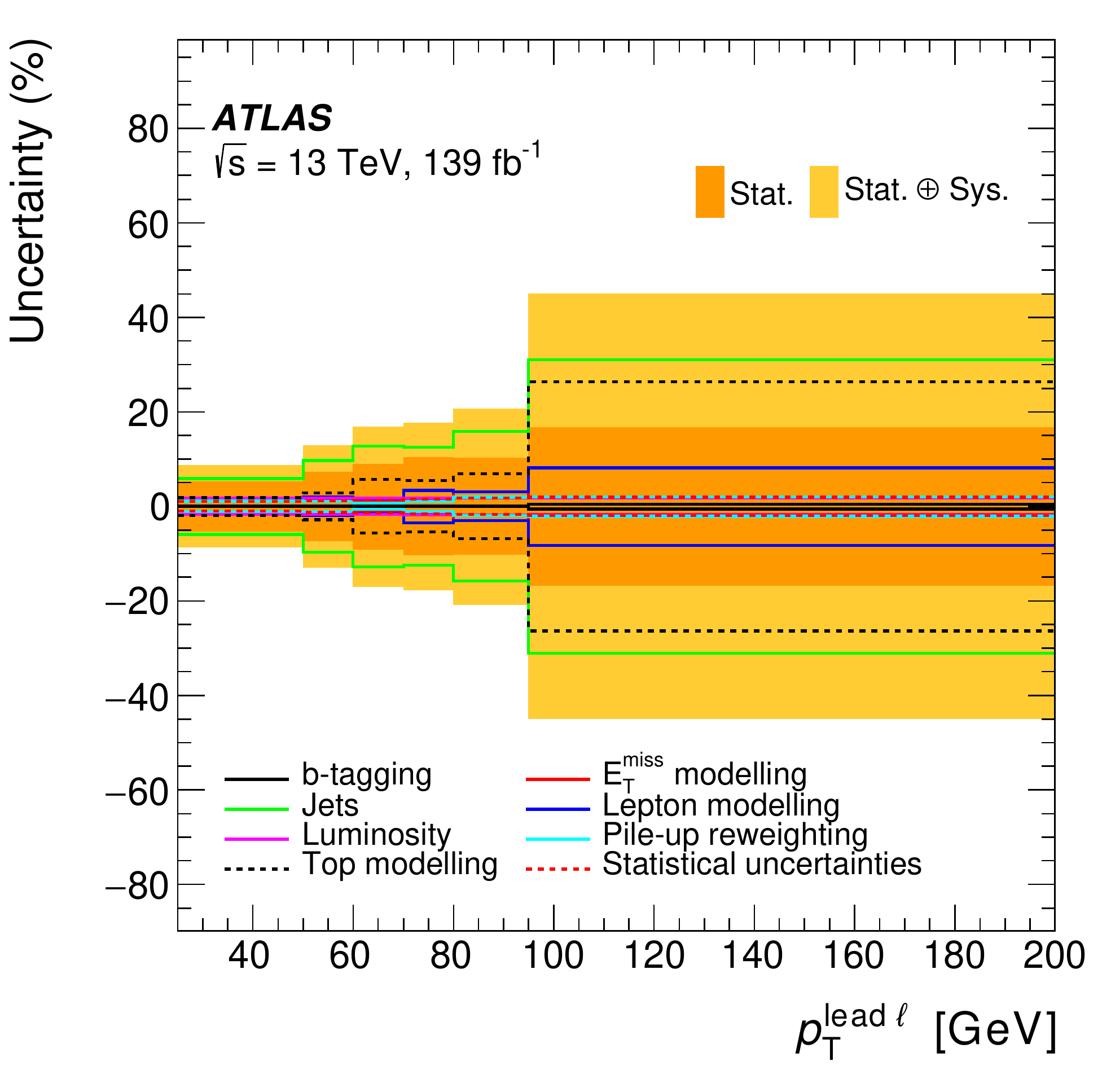} \\
\includegraphics[width=0.41\linewidth]{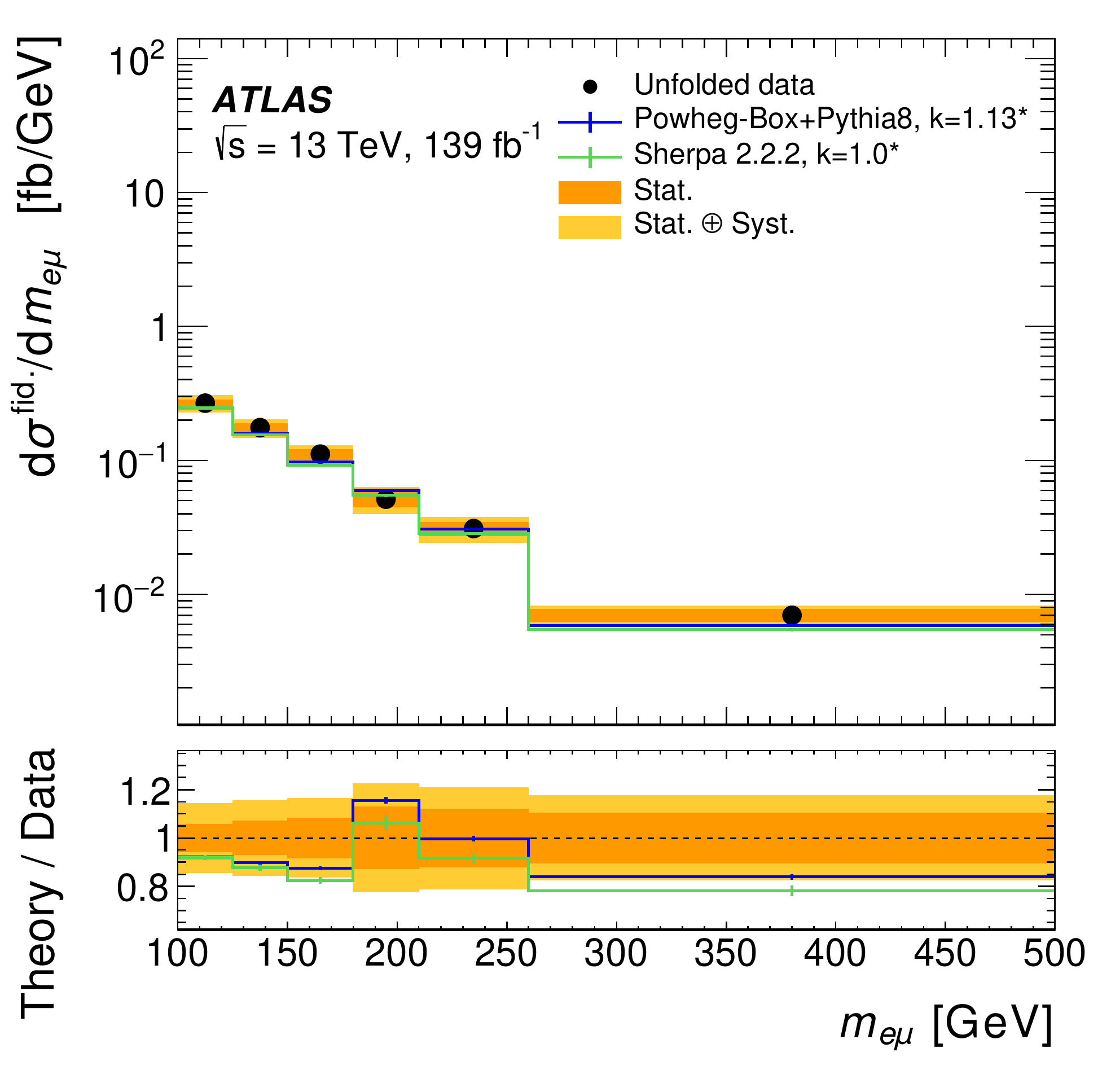}
\includegraphics[width=0.41\linewidth]{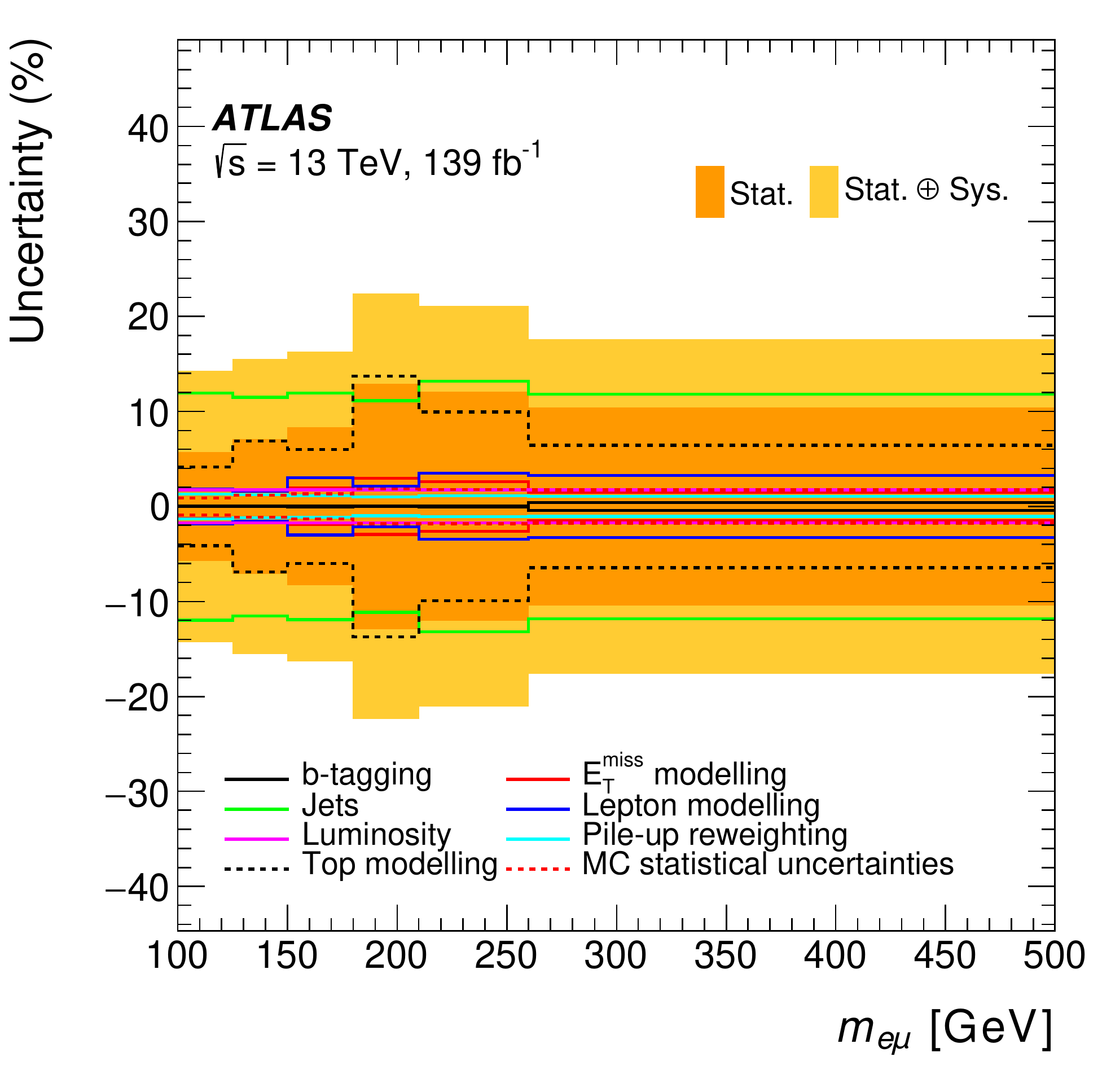} \\
 
\includegraphics[width=0.41\linewidth]{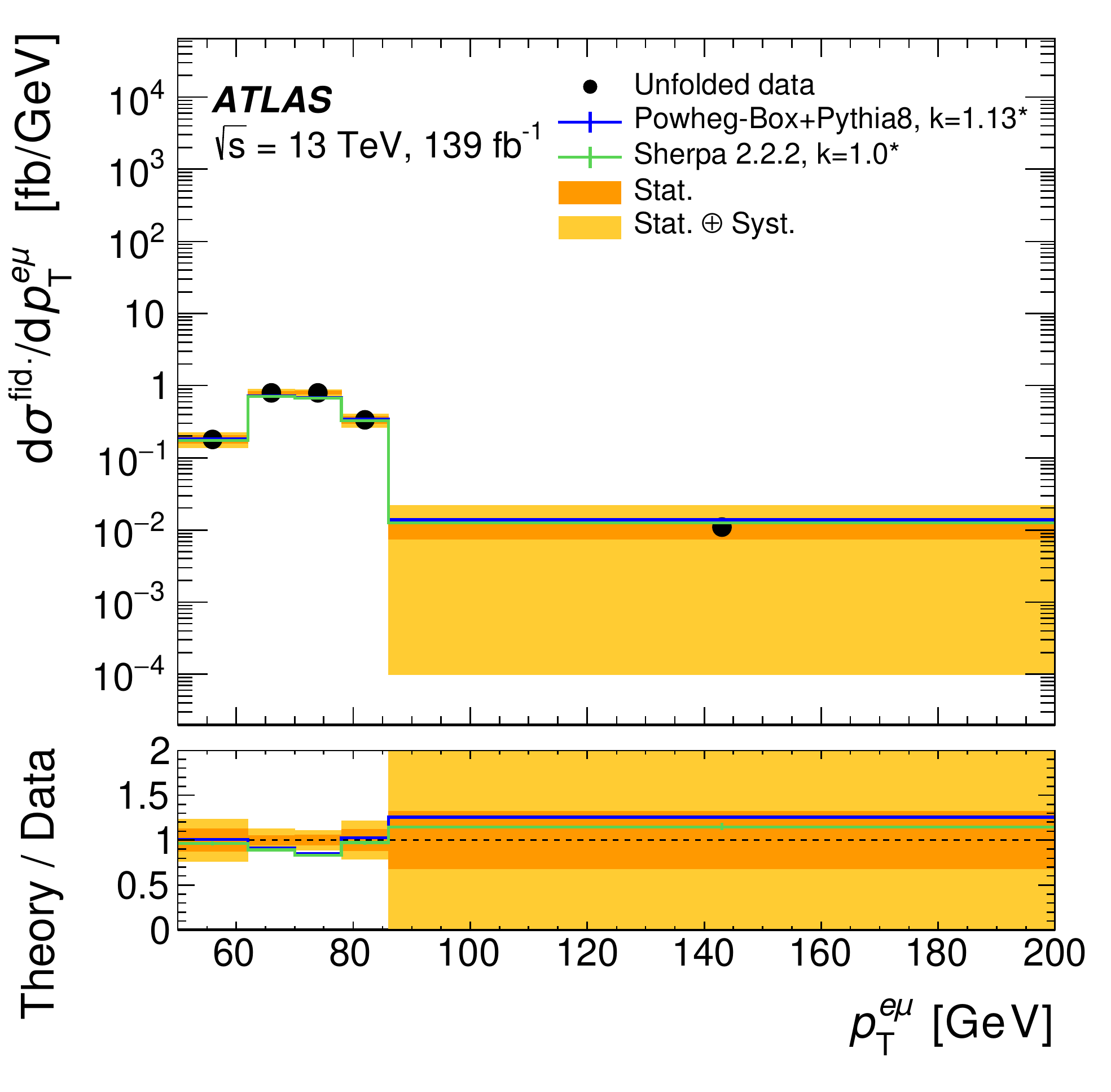}
\includegraphics[width=0.41\linewidth]{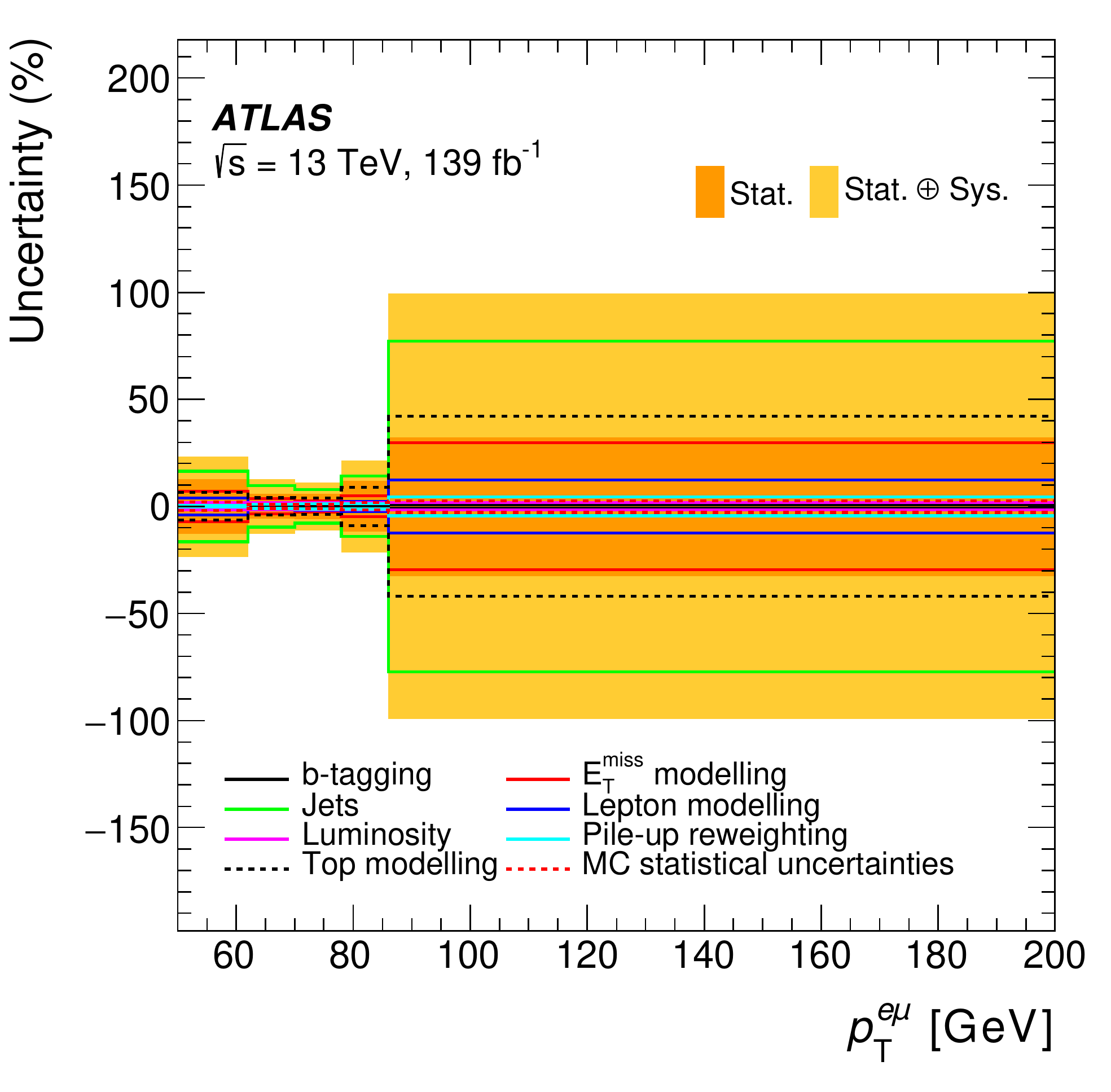} \\
 
\caption{Measured fiducial differential cross-sections of $WW$ production for (top to bottom)   $p_{\mathrm{T}}^{\mathrm{lead}\, \ell}$, $m_{e\mu}$ and $p_{\mathrm{T}}^{e\mu}$. The last bin is inclusive in the measured observable and for $p_{\mathrm{T}}^{e\mu}$ the first bin contains the underflow bin. The measured cross-section values are shown as points with dark bands giving the statistical uncertainty and light bands indicating the size of the total uncertainty. The results are compared with the $q\bar{q}$-initiated predictions from \POWHEGBOX[v2] +\,\PYTHIA[8.186] and \SHERPA[2.2.2], each combined with  \SHERPA[2.2.2] +\,\OPENLOOPS (LO+PS) for the $gg$-initiated states. The $k$-factors refer to the corrections applied to scale the predictions of $qq$-initiated and $gg$-initiated processes to NNLO and NLO accuracy in QCD respectively. The right column shows a breakdown of contributions to the uncertainties in the unfolded measurement.  }
\label{fig:unfoldedresultsScale}
\end{figure}

Particle-level differential cross-sections for the six variables targeted in this study are presented in Figure~\ref{fig:unfoldedresultsAng} for the angular variables and Figure~\ref{fig:unfoldedresultsScale} for the scale variables. In each case, the right-hand plot shows the impact of the uncertainties, grouped into the categories discussed in Section~\ref{sec:systematics}, on the measurement.
 
The dilepton rapidity distribution has a maximum between $|y_{e\mu}|=0$ and $|y_{e\mu}|=1$, consistent with central production of a massive diboson system. The $|\Delta\phi_{e\mu}|$ distribution peaks at $|\Delta\phi_{e\mu}|=2$. The shape of this distribution is influenced by the \mttwo\ selection defining the fiducial region: high $|\Delta\phi_{e\mu}|$ values are associated with back-to-back leptons, which typically give lower \mttwo\ values than are considered in this study. Conversely, the highest \mttwo\ values (which for $WW$ production should occur around 90~\GeV\ in the absence of detector effects) are often associated with collinear leptons (low $|\Delta\phi_{e\mu}|$), which are also excluded from the fiducial region. The  $\cos\theta^{\ast}$ distribution peaks around $\cos\theta^{\ast}=0.8$, with higher values being suppressed by the rapidity acceptance of the fiducial phase space. The distributions of the scale variables all show the expected characteristic fall for high values of the variable. The fiducial phase-space acceptance also suppresses the leading-lepton and dilepton \pT\ distributions at lower \pT\ values.
 
The measurements are compared with the $qq$-initiated NLO QCD+PS predictions from \POWHEGBOX[v2] +\,\PYTHIA[8.186] and \SHERPA[2.2.2], each combined with  \SHERPA[2.2.2] +\,\OPENLOOPS (LO+PS) for the $gg$-initiated states.  For the angular variables, the region with $|\Delta\phi_{e\mu}|<1.5$ is underestimated by both theory predictions, which is consistent with observations in the previous ATLAS 13~\TeV\ $WW$+0-jet measurement~\cite{STDM-2017-24}. The region  $\cos\theta^{\ast}>0.8$ is also underestimated by 10\%--30\% by both predictions. This corresponds to a rapidity difference of $|\Delta y(e\mu)|\geq2.2$ between the leptons. For the  distribution of dilepton rapidity $|y_{e\mu}|$, the theory shows reasonable agreement with the measurement.  The predictions for the scale variables show good agreement with the data except for low values of $p_{\mathrm{T}}^{\mathrm{lead}\, \ell}$, where both predictions underestimate the cross-section by 20\%--25\%. Global $\chi^{2}$ calculations are carried out for all predictions and are displayed in Table~\ref{tab:chisquared}. Uncertainties in the theory predictions are not considered. The largest $\chi^{2}/\textup{NDF}$ is 24.1 / 6 corresponding to the comparison between the $q\bar{q}\rightarrow WW $ (\SHERPA[2.2.2]) +\,$gg\rightarrow WW$ (\SHERPA[2.2.2] +\,\OPENLOOPS) prediction and the unfolded distribution for $p_{\mathrm{T}}^{\mathrm{lead}\, \ell}$.

\FloatBarrier
 
\begin{table}[htb!]
\caption{Chi-squared per number of degrees of freedom $\chi^{2}/\textup{NDF}$ for a comparison of unfolded distributions with different theory predictions. The  calculation takes into account bin-by-bin correlations of systematic and statistical uncertainties. Uncertainties in the theory predictions are not considered. }
\centering
\begin{tabular}{l|cccccc}
\hline
 
&  $ |y_{e\mu}|$ & $ |\Delta\phi_{e\mu}|$ & $ \cos\theta^{\ast}$ & $p_{\mathrm{T}}^{\mathrm{lead}\,\ell}$ & $m_{e\mu}$ & $p_{\mathrm{T}}^{e\mu}$ \\
\hline
~~~~\POWHEGBOX[v2] + \PYTHIA[8] ($q\bar{q}$)  & \multirow{2}{*}{14.4 / 8 }  &  \multirow{2}{*}{10.1 / 10}     &  \multirow{2}{*}{13.3 / 7}   &  \multirow{2}{*}{15.4 / 6  }       &  \multirow{2}{*}{2.8 / 6 }   &  \multirow{2}{*}{3.9 / 5 } \\
\& \Sherpa[2.2.2] + \OPENLOOPS ($gg$)   \\
\hline
~~~~\Sherpa[2.2.2] ($q\bar{q}$)    & \multirow{2}{*}{18.3 / 8 }   &	\multirow{2}{*}{17.9 / 10} &	\multirow{2}{*}{24.5 / 7}  & 	\multirow{2}{*}{24.1 / 6 }        &  \multirow{2}{*}{2.5 / 6}    & \multirow{2}{*}{4.1 / 5 }  \\
\& \Sherpa[2.2.2] + \OPENLOOPS ($gg$)   \\
\hline
\end{tabular}
\label{tab:chisquared}
\end{table}
 
\FloatBarrier

\section{Conclusion}
\label{sec:conclusions}
 
The  cross-section for $WW \rightarrow e^{\pm}\nu\mu^{\mp}\nu$ production in $pp$ collisions at $\sqrt{s}=13$~\TeV\ is measured with the ATLAS detector at the LHC in a fiducial phase-space  characterised by the absence of jets and additional leptons, the presence of a high dilepton invariant mass $m_{e\mu}$, and with values of \MET\ and the stransverse mass \mttwo\ motivated by the control regions used in supersymmetry searches~\cite{SUSY-2018-32}. The measured cross-section is $\sigma(WW \rightarrow e^{\pm}\nu\mu^{\mp}\nu)= 19.2 \pm 0.3~(\mathrm{stat}) \pm 2.5~(\mathrm{syst}) \pm 0.4~(\mathrm{lumi})~\mathrm{fb}$. Differential cross-sections for three variables sensitive to the energy scale of the event and three variables sensitive to the angular correlations of the leptonic decay products are compared with two theoretical SM predictions from perturbative QCD calculations. Good agreement is observed for most distributions within the uncertainties. The largest discrepancies occur at low values of $|\Delta\phi_{e\mu}|<1.5$, high values of $\cos\theta^{\ast}>0.8$ and low $p_{\mathrm{T}}^{\mathrm{lead}\, \ell}$, which is consistent with the observations of the previous ATLAS $WW$+0-jet measurement~\cite{STDM-2017-24}.
This study validates the SM in a new and interesting region motivated particularly by searches for supersymmetry and provides benchmark measurements that can be used to improve future SM predictions and calculate additional constraints on BSM models.

\section*{Acknowledgements}

% The next lines are included from the .//acknowledgements/Acknowledgements.tex input file

We thank CERN for the very successful operation of the LHC, as well as the
support staff from our institutions without whom ATLAS could not be
operated efficiently.
 
We acknowledge the support of
ANPCyT, Argentina;
YerPhI, Armenia;
ARC, Australia;
BMWFW and FWF, Austria;
ANAS, Azerbaijan;
CNPq and FAPESP, Brazil;
NSERC, NRC and CFI, Canada;
CERN;
ANID, Chile;
CAS, MOST and NSFC, China;
Minciencias, Colombia;
MEYS CR, Czech Republic;
DNRF and DNSRC, Denmark;
IN2P3-CNRS and CEA-DRF/IRFU, France;
SRNSFG, Georgia;
BMBF, HGF and MPG, Germany;
GSRI, Greece;
RGC and Hong Kong SAR, China;
ISF and Benoziyo Center, Israel;
INFN, Italy;
MEXT and JSPS, Japan;
CNRST, Morocco;
NWO, Netherlands;
RCN, Norway;
MEiN, Poland;
FCT, Portugal;
MNE/IFA, Romania;
MESTD, Serbia;
MSSR, Slovakia;
ARRS and MIZ\v{S}, Slovenia;
DSI/NRF, South Africa;
MICINN, Spain;
SRC and Wallenberg Foundation, Sweden;
SERI, SNSF and Cantons of Bern and Geneva, Switzerland;
MOST, Taiwan;
TENMAK, T\"urkiye;
STFC, United Kingdom;
DOE and NSF, United States of America.
In addition, individual groups and members have received support from
BCKDF, CANARIE, Compute Canada and CRC, Canada;
PRIMUS 21/SCI/017 and UNCE SCI/013, Czech Republic;
COST, ERC, ERDF, Horizon 2020 and Marie Sk{\l}odowska-Curie Actions, European Union;
Investissements d'Avenir Labex, Investissements d'Avenir Idex and ANR, France;
DFG and AvH Foundation, Germany;
Herakleitos, Thales and Aristeia programmes co-financed by EU-ESF and the Greek NSRF, Greece;
BSF-NSF and MINERVA, Israel;
Norwegian Financial Mechanism 2014-2021, Norway;
NCN and NAWA, Poland;
La Caixa Banking Foundation, CERCA Programme Generalitat de Catalunya and PROMETEO and GenT Programmes Generalitat Valenciana, Spain;
G\"{o}ran Gustafssons Stiftelse, Sweden;
The Royal Society and Leverhulme Trust, United Kingdom.
 
The crucial computing support from all WLCG partners is acknowledged gratefully, in particular from CERN, the ATLAS Tier-1 facilities at TRIUMF (Canada), NDGF (Denmark, Norway, Sweden), CC-IN2P3 (France), KIT/GridKA (Germany), INFN-CNAF (Italy), NL-T1 (Netherlands), PIC (Spain), ASGC (Taiwan), RAL (UK) and BNL (USA), the Tier-2 facilities worldwide and large non-WLCG resource providers. Major contributors of computing resources are listed in Ref.~\cite{ATL-SOFT-PUB-2021-003}.

% End of text imported from the .//acknowledgements/Acknowledgements.tex input file

\clearpage

\printbibliography

\clearpage
% ATLAS Collaboration author list
% Reference date of SUSY-2021-06 is 2022-02-08
% Author list last updated on date 09-NOV-23
% Data extracted on 09-Nov-2023 for paper reference SUSY-2021-06
% at 5:49pm
 
\begin{flushleft}
\hypersetup{urlcolor=black}
{\Large The ATLAS Collaboration}

\bigskip

\AtlasOrcid[0000-0002-6665-4934]{G.~Aad}$^\textrm{\scriptsize 101}$,
\AtlasOrcid[0000-0002-5888-2734]{B.~Abbott}$^\textrm{\scriptsize 119}$,
\AtlasOrcid[0000-0002-7248-3203]{D.C.~Abbott}$^\textrm{\scriptsize 102}$,
\AtlasOrcid[0000-0002-1002-1652]{K.~Abeling}$^\textrm{\scriptsize 55}$,
\AtlasOrcid[0000-0002-8496-9294]{S.H.~Abidi}$^\textrm{\scriptsize 29}$,
\AtlasOrcid[0000-0002-9987-2292]{A.~Aboulhorma}$^\textrm{\scriptsize 35e}$,
\AtlasOrcid[0000-0001-5329-6640]{H.~Abramowicz}$^\textrm{\scriptsize 150}$,
\AtlasOrcid[0000-0002-1599-2896]{H.~Abreu}$^\textrm{\scriptsize 149}$,
\AtlasOrcid[0000-0003-0403-3697]{Y.~Abulaiti}$^\textrm{\scriptsize 116}$,
\AtlasOrcid[0000-0003-0762-7204]{A.C.~Abusleme~Hoffman}$^\textrm{\scriptsize 136a}$
% ...
\bigskip
\\

% ...
$^{29}$Physics Department, Brookhaven National Laboratory, Upton NY; United States of America.\\
% ...
$^{35}$$^{(a)}$Facult\'e des Sciences Ain Chock, R\'eseau Universitaire de Physique des Hautes Energies - Universit\'e Hassan II, Casablanca;$^{(b)}$Facult\'{e} des Sciences, Universit\'{e} Ibn-Tofail, K\'{e}nitra;$^{(c)}$Facult\'e des Sciences Semlalia, Universit\'e Cadi Ayyad, LPHEA-Marrakech;$^{(d)}$LPMR, Facult\'e des Sciences, Universit\'e Mohamed Premier, Oujda;$^{(e)}$Facult\'e des sciences, Universit\'e Mohammed V, Rabat;$^{(f)}$Institute of Applied Physics, Mohammed VI Polytechnic University, Ben Guerir; Morocco.\\
% ...
$^{55}$II. Physikalisches Institut, Georg-August-Universit\"{a}t G\"ottingen, G\"ottingen; Germany.\\
% ...
$^{101}$CPPM, Aix-Marseille Universit\'e, CNRS/IN2P3, Marseille; France.\\
$^{102}$Department of Physics, University of Massachusetts, Amherst MA; United States of America.\\
% ...
$^{116}$Department of Physics, New York University, New York NY; United States of America.\\
% ...
$^{119}$Homer L. Dodge Department of Physics and Astronomy, University of Oklahoma, Norman OK; United States of America.\\
% ...
$^{136}$$^{(a)}$Departamento de F\'isica, Pontificia Universidad Cat\'olica de Chile, Santiago;$^{(b)}$Millennium Institute for Subatomic physics at high energy frontier (SAPHIR), Santiago;$^{(c)}$Instituto de Investigaci\'on Multidisciplinario en Ciencia y Tecnolog\'ia, y Departamento de F\'isica, Universidad de La Serena;$^{(d)}$Universidad Andres Bello, Department of Physics, Santiago;$^{(e)}$Instituto de Alta Investigaci\'on, Universidad de Tarapac\'a, Arica;$^{(f)}$Departamento de F\'isica, Universidad T\'ecnica Federico Santa Mar\'ia, Valpara\'iso; Chile.\\
% ...
$^{149}$Department of Physics, Technion, Israel Institute of Technology, Haifa; Israel.\\
$^{150}$Raymond and Beverly Sackler School of Physics and Astronomy, Tel Aviv University, Tel Aviv; Israel.\\
% ...
$^{a}$ Also Affiliated with an institute covered by a cooperation agreement with CERN.\\
$^{b}$ Also at Borough of Manhattan Community College, City University of New York, New York NY; United States of America.\\
$^{c}$ Also at Bruno Kessler Foundation, Trento; Italy.\\
$^{d}$ Also at Center for High Energy Physics, Peking University; China.\\
$^{e}$ Also at Centro Studi e Ricerche Enrico Fermi; Italy.\\
$^{f}$ Also at CERN, Geneva; Switzerland.\\
% ...

\end{flushleft}

% Created with Glance <Atlas.Glance@cern.ch>

\clearpage

\FloatBarrier

\end{document}